# Four-band non-Abelian topological insulator and its experimental realization

Tianshu Jiang[1†], Qinghua Guo[1†], Ruo-Yang Zhang[1], Zhao-Qing Zhang[1], Biao Yang[1,2*], C. T. Chan[1*]

[1]*Department of Physics and Institute for Advanced Study, The Hong Kong University of Science and Technology, Hong Kong, China*

[2]*College of Advanced Interdisciplinary Studies, National University of Defense Technology, Changsha 410073, China*

[†]These authors contributed equally to this work.

[*]Correspondence to: yangbiaocam@nudt.edu.cn; phchan@ust.hk;

**Abstract**

**Very recently, increasing attention has been focused on non-Abelian topological charges, e.g. the quaternion group $Q_8$. Different from Abelian topological band insulators, these systems involve multiple tangled bulk bandgaps and support non-trivial edge states that manifest the non-Abelian topological features. Furthermore, a system with even or odd number of bands will exhibit significant difference in non-Abelian topological classifications. Up to now, there is scant research investigating the even-band non-Abelian topological insulators. Here, we both theoretically explored and experimentally realized a four-band PT (inversion and time-reversal) symmetric system, where two new classes of topological charges as well as edge states are comprehensively studied. We illustrate their difference from four-dimensional rotation senses on the stereographically projected Clifford tori. We show the evolution of bulk topology by extending the 1D Hamiltonian onto a 2D plane and provide the accompanying edge state distributions following an analytical method. Our work presents an exhaustive study of four-band non-Abelian topological insulators and paves the way to other even band systems.**

In mathematics, Abelian operators are commutative meaning that the result of two successive operations does not depend on the order in which they are written. If we focus on a single bandgap, topological physical systems[1-6] are usually classified by Abelian groups, with the prime example being the ten-fold classification[7,8] of Hermitian topological insulators and superconductors. Once multiple bandgaps are collectively considered, their coupling introduces richer physics that can make the classification non-Abelian[9-13]. A classic example is the quaternion group $Q_8 = \{+1, \pm i, \pm j, \pm k, -1\}$ with $i^2 = j^2 = k^2 = ijk = -1$, which has been used to classify the topological line defects in biaxial nematic liquid crystals[14]. Very recently, the non-Abelian group was used to describe the admissible nodal line configurations[12,15,16], Dirac/Weyl point braiding[13,17,18] and intriguing triple nodal points[19-21] in PT (inversion and time-reversal) symmetric systems. When more bands are involved, richer non-Abelian topological charges emerge[9]. Especially for the systems with even number of bands, several new classes of non-Abelian topological charges deserve special attention. A simple argument to this is that the even dimensional special orthogonal groups, i.e. $SO(2N)$ with $N$ indicating positive integer, contain inversion symmetry, i.e. $-I_{2N}$ (the negative $2N \times 2N$ identity matrix).

**Non-Abelian topological charges in four-band models**

Here, for simplicity we focus on a four-band PT symmetric system. Choosing an appropriate basis the Hamiltonian can take real forms, i.e. $H(k) = H^*(k)$ . When simultaneously considering all the three bandgaps between any two adjacent bands, the configuration space of the Hamiltonian is $M_4 = O(4)/\mathbb{Z}_2^4$, with $O(4)$ being the 4D orthogonal group. This implies that the eigenstate frame remains intact under $O(4)$ rotation, while $\mathbb{Z}_2^4$ indicates each eigenstate has the gauge freedom of $\pm 1$. The quantized charges that describe the underlying topology are found to be the non-Abelian based homotopy group[9] $\pi_1(M_4) = Q_{16} = \cup_{n_i \in \{0,1\}}\{\pm e_1^{n_1} e_2^{n_2} e_3^{n_3}\}$, where $e_1, e_2, e_3$ are the basis of real Clifford algebra $C\ell_{0,3}$ with satisfying the relation $\{e_i, e_j\} = -2\delta_{ij}$. There are 16 elements in the group and 10 conjugacy classes in total (see Table 1 as indicated by the curly braces). The group multiplication can be simply carried out with using the above relation, i.e. $(e_1 e_2)(e_1 e_3) = -e_1 e_1 e_2 e_3 = e_2 e_3$. Although the labels with the Clifford algebra basis (see the 1st column of Table 1) are convenient for group multiplication, it is not straightforward to decode the underlying physical meaning. In order to relate the charges to the rotations of the eigenstates, we rename all the charges one-to-one as shown in the 2nd column of Table 1. For example, we will see that $\pm q_{12}$ indicate that both the 1st and 2nd bands acquire Zak phases of $\pi$ due to the rotation of their respective eigenvectors. Figure 1a shows the representative elements and their multiplication relations, the corresponding full multiplications are listed in Supplementary Tables 1 and 2. One may also notice that the paths (arrows) bridging two elements are not unique. It means the non-Abelian topological phase transitions are multiple-pathed, which is different from the single-pathed way in Abelian systems[22].

In the following we study the topological properties of those charges. After topological band flattening, the mentioned PT symmetric four-band Hamiltonian can take the form of $H(k) = R(k)I_{1234}(k)R^T$ with $R(k) \in SO(4)$ being the 4D special orthogonal group, $k \in [-\pi, \pi]$ being the first Brillouin zone (FBZ) and $I_{1234} = diag(1,2,3,4)$. The Hamiltonian has four real eigenvectors as $H(k)|n\rangle = n|n\rangle$ with $n$ =1, 2, 3 and 4. When $k$ runs across the FBZ ($k = -\pi \rightarrow \pi$), the rotation matrix $R(k)$ will continuously act on the eigenvector $|n\rangle$ and one finally obtains $+ \ or \ - |n\rangle$ corresponding to the Zak phase of $0 \ or \ \pi$, respectively. Without loss of generality, we assume $R(k = -\pi) = I_4$. Because $\det(R) = \lambda_1\lambda_2\lambda_3\lambda_4 = 1$ with $\lambda_i$ being the four eigenvalues of $R(k)$, it is easy to find there are exhaustively three categories of possibilities at $k = \pi$: (1) All four $\lambda_i = 1$; (2) Two $\lambda_i = 1$, the other two $\lambda_i = -1$; (3) All four $\lambda_i = -1$.

The first category corresponds to two conjugacy classes $\{+1\}$ and $\{-1\}$. Although they are indistinguishable from the Zak phase description, charge $+1$ indicates the trajectories of eigenstate frame are contractible, while charge $-1$ indicates an non-contractible loop. Usually, charge $-1$ indicates that the eigenstate frame rotates $2\pi$ in a rotation plane (or topologically equivalent configurations)[9,22]. We will see their difference more explicitly by extending the 1D Hamiltonian onto a 2D plane (Figs. 1d and e); The second category consists of six conjugacy classes which can be distinguished using single band Zak phase arguments, regarding which two of the four bands have Zak phases of $\pi$; In the last category, all eigenstates flip their sign after $k$ runs across the 1D FBZ. This category originates from the inversion symmetry ($-I_4$) mentioned above. The two group elements (classes) also share the same Zak phase distribution and are indistinguishable from the conventional Abelian arguments. Their difference is reflected in the sense of eigenstate rotation in four-dimension.

With setting $R(k) = \exp\left(\phi \sum_{i<j=1:4} n_{ij}L_{ij}\right)$, we obtain the explicit form of the flat band Hamiltonian, where six skew-symmetric matrices $L_{ij}$ with entries $\left(L_{ij}\right)_{a,b=1:4} = -\delta_{ia}\delta_{jb} + \delta_{ib}\delta_{ja}$ span the basis of Lie algebra $\mathfrak{so}(4)$, $\phi(k)$ is the rotation angle and $n_{ij}(k)$ determines the rotation plane. For example, the Hamiltonian of charge $q_{12}$ can be given with $R(k) = \exp\left(\frac{k+\pi}{2}L_{12}\right)$, while charge $-1$ can be obtained with $R(k) = \exp[(k+\pi)L_{12}]$. Except for the charges of $\pm q_{1234}$, the rest have their counterparts in the three-band systems[22] studied previously. Thus, we mainly focus on the charges $\pm q_{1234}$ which are unique in the four-band models.

While the non-Abelian topological charges are defined on one-dimensional periodic lattices, their topological characters would be more straightforward to visualize after we generalize the 1D Hamiltonians onto a 2D extended plane, where each non-Abelian topological charge characterizing the 1D loop is reflected by the specific configuration of band degeneracies encircled by the 1D loop in the 2D plane. After trigonometrically expanding the Hamiltonian $H(k)$, we make the substitutions like $\cos k \to \rho \cos k = k_1$ and $\sin k \to \rho \sin k = k_2$ and show the corresponding two-dimensional bands in Figs. 1b-e. The original 1D Hamiltonian in $k$ space is a unit circle (white circles in Figs. 1b-e) in the 2D extended plane which encircles non-removable degeneracies explicitly exhibiting the underlying topological obstacles. The charge $+1$ (Fig. 1e), it is trivial as there is no degeneracy enclosed by the white circles, while for charges $\pm q_{mn}$ (Fig. 1b) and $-1$ (Fig. 1d) the 1D unit circles enclose linear and quadratic degeneracies, respectively. These 2D degeneracies will topologically contribute to edge/domain-wall states of the 1D systems, i.e. the linear/quadratic degeneracy implies one/two topologically protected edge states.

The charge $q_{1234}$ can be factorized as $q_{1234} = q_{12}q_{34}$, $q_{1234} = q_{14}q_{23}$ and $q_{1234} = -q_{13}q_{24}$ (the minus sign is induced by the odd permutation of subscripts). Note that the two factors are commutative, i.e. $q_{12}q_{34} = q_{34}q_{12}$, in nodal links which means all nodes formed by more distant (i.e. sharing no common band) pairs of bands commute[9]. In Fig. 1c we show the corresponding extended 2D band degeneracies of the three cases. They all belong to the same charge and thus can be continuously transformed into each other without closing the bandgap (see Figure 3). The charge $-q_{1234}$ shares the same two-dimensional band degeneracies with $q_{1234}$. Note that $\pm q_{1234}$ belong to two different conjugacy classes, which is one of the key points that fundamentally distinguishes them from the charges $\pm q_{mn}$. We will show their topological differences in the following section from the eigenstate rotation perspective.

We note that the nodal ring degeneracies in Figs. 1b ($\pm q_{14}$) and c ($-q_{13}q_{24}$) are accidental in the flat band models, each will be split into linear Dirac cones in more general situations (see Figure 5b and Supplementary Figure 16c, respectively). Other triple degeneracies are similar to charges $\pm j$ in three-band models[22], where there are three bands being involved. The four-fold degeneracy in Fig. 1c ($q_{14}q_{23}$) is also admissible rather than stable here.

**Eigenstates on three-sphere - $S^3$**

Here, we illustrate rotation configurations pertaining to different charges of the generalized quaternion group $Q_{16}$. The normalized eigenstates of $H(k)$ are all real and can be parametrized by Hopf coordinates $(\alpha, \eta, \beta)$ on the three-sphere - $S^3$. Their four components can be written as $(u = \cos\alpha \sin\eta\,, x = \sin\alpha \sin\eta\,, y = \cos\beta \cos\eta\,, z = \sin\beta \cos\eta)$, where $\alpha$ and $\beta$ respectively correspond to the two rotation angles in the two orthogonal invariant planes as shown in Fig. 2a (also see Supplementary Information II: Rotations in four-dimension[23-25]), while $\eta$ determines the proportions projected onto the two planes. When $\alpha \neq 0, \beta = 0$ (or $\alpha = 0, \beta \neq 0$), the rotations are called single rotations. For example, all ideal rotations $R(k)$ with $k = -\pi \to \pi$ enabling charge $q_{12}$ belong to the case with setting $\eta = \frac{\pi}{2}$ and $\alpha = \frac{k+\pi}{2}$, where "ideal" indicates the flat band model given above. Note that all general models can be continuously transformed into the ideal flat band model and they are topologically equivalent. Other charges including $\pm q_{mn}$ and $-1$ can be realized in a similar manner. It is easy to see that the eigenstates in one plane (i.e. $oyz$ plane when $\eta = \frac{\pi}{2}$) can be fixed for these cases, while they rotate on the other orthogonal plane (i.e. $oux$ plane). In other words, the ideal rotations can be carried out in a two-dimensional subspace. It is worth pointing out that in contrast on which plane the eigenstates rotate, the crucial property of these topological charges is that the trajectories of eigenstates cannot contract to isolated points. The difference of charges $\pm q_{mn}$ is reflected by which two bands (the mth and nth) are noncontractible, while charge $-1$ requires all four trajectories cannot contract simultaneously.

When both $\alpha \neq 0$ and $\beta \neq 0$, the rotations are dubbed as double rotations (Fig. 2a), where there are two possibilities: rotating on the two planes in the same ($\alpha\beta > 0$) or opposite ($\alpha\beta < 0$) senses. The charges $\pm q_{1234}$ have to be realized with continuous double rotations, which means $R(k)$ at each $k$ point is a double rotation. Interestingly, when $\eta = \frac{\pi}{4}$, the parametric set $(u, x, y, z)$ constructs a Clifford torus[26], which is the Cartesian product of two circles in $\mathbb{R}^4$ (e.g. $S_A^1 \in oux, S_B^1 \in oyz\ and\ S_A^1 \times S_B^1 \in \mathbb{R}^4$). The Clifford torus can be stereographically projected[26] into $\mathbb{R}^3$ as a conventional torus, i.e. $\left(\frac{x}{1-u}, \frac{y}{1-u}, \frac{z}{1-u}\right)$, on which we can pictorially illustrate the difference between charges $\pm q_{1234}$ from the rotation senses of eigenstate trajectories as shown in Fig. 2b. The two panels correspond to $\alpha = \beta = \frac{k+\pi}{2}$ (left, $q_{1234}$) and $\alpha = -\beta = \frac{k+\pi}{2}$ (right, $-q_{1234}$), respectively (see other cases in Supplementary Figure 5).

We further propose another orthographical projection method, which projects each 4D trajectory into 3D space from four orthogonal views. This is similar to the three-view drawing, which is the orthographic projection form 3D space to 2D plane. Take the first panel of Fig. 2c as an example, we plot the trajectories in the $xyz$ subspace so that it is an orthographic projection from the view of $u$ direction. Figures 2c and d respectively correspond to $+q_{1234}$ and $-q_{1234}$, where eigenstate trajectories are mapped onto four solid spheres in $\mathbb{R}^3$. One can see their only difference is that the rotation directions in the $oux$ plane are opposite. Orthographical projections for other charges are listed in Supplementary Figures 1-4. In Fig. 2e, we show the topological phase transition between them, where there are inevitably two linear crossings between the first and second bands as system parameter $w_{AB}$ changes (without relying on a joint basepoint as they belong to different classes).

**Zak phases and evolution of edge states**

After understanding the non-Abelian topological charges from the perspective of eigenstate frame rotations, we now show their relations to the Zak phases of each band as well as edge/domain-wall states. In a PT-symmetric system, the Zak phases of each band will take the quantized values of 0 or $\pi$, which have been shown in Table 1, i.e. $\lambda_i = -1$ indicates Zak phase of $\pi$. We further refine the Zak phase of $\pi$ to be $\pm\pi$, where "$\pm$" is used to differentiate between charges $\pm q_{mn}$ (two elements in the same conjugacy class). All of the corresponding single band Zak phases are exhaustively summarized in Fig. 3a. For charges $\pm q_{mn}$, two corresponding bands with noncontractible eigenstate trajectories carry Zak phases of $\pm\pi$, and the bandgap sandwiched by them supports edge states at hard boundaries of a finite lattice. We take the case of $\pm q_{12}$ as an example as shown in Fig. 3b. Edge states of other $\pm q_{mn}$ charges are shown in the Supplementary Figure 6. We label charge $-1$ with $2\pi$, which indicates noncontractible $2\pi$ rotation here[22].

For charges $\pm q_{1234}$, two eigenstates rotate $\pi$, while the other two rotate $\pm\pi$ when $k = -\pi \rightarrow \pi$, respectively. As shown in Fig. 1c, there are three ways of factorization. We further show them schematically in Fig. 3c, where each double-headed arrow represents one factorization. The commutative property between two factor charges, i.e. $q_{12}q_{34} = q_{34}q_{12}$, is implied by the double-headed arrows. The fact that $q_{12}q_{34}$ (type-I), $-q_{13}q_{24}$ (type-II) and $q_{14}q_{23}$ (type-III) are the same element in the group can be visualized by constructing a transformation between them without gap closing. The continuous transition between different factorizations can be explicitly parameterized. For example from $q_{12}q_{34} \rightarrow -q_{13}q_{24}$, we have $H(k) = R_2 R_1 I_{1234} R_1^{-1} R_2^{-1}$, with $R_1(k) = \exp[(k+\pi)/2\,(\cos\theta_{I\rightarrow II}\,L_{12} - \sin\theta_{I\rightarrow II}\,L_{13})]$ and $R_2(k) = \exp[(k+\pi)/2\,(\cos\theta_{I\rightarrow II}\,L_{34} + \sin\theta_{I\rightarrow II}\,L_{24})]$ as shown in Fig. 3c (see the evolution of eigenstate trajectories in Supplementary Figure 3). In other words, the pair of two orthogonal invariant planes rotates with $\theta_{I\rightarrow II}$. We further study the accompanying evolution of edge states at hard boundaries as shown in Figs. 3d-f. They have analytical results: $E^{\pm} = \frac{5}{2} \pm \frac{\sqrt{2}}{4}\sqrt{5 + 3\cos 2\theta_{I\rightarrow II}}$, $E^{\pm} = \frac{5}{2} \pm \frac{1}{2}\cos\theta_{II\rightarrow III}$ and $E^{\pm} = \frac{5}{2} \pm \sin\theta_{III\rightarrow I}$, respectively. Detailed analytical methods are provided in Supplementary Information IV: Analytical solutions of edge states for the flat-band models. There are a total of two edge states pumping between different bandgaps. Their field distributions are given in the Supplementary Figures 7-9. The existence of these edge modes can be inferred heuristically by examining the band degeneracies of the extended 2D model. In Figs. 3g-i, we show the radial cuts of their extended two-dimensional bands, where one can easily finds that each linear degeneracy point at $k_r = 0$ implies the position of each edge state in Figs. 3d-f, respectively. Note that in these flat band cases, only the degeneracies at $k_r = 0$ imply topological edge states, while other degeneracies ($k_r \neq 0$) accidentally emerge from the 2D nodal rings (e.g. see Fig. 1c) which have no topological implication.

We also show the edge state evolution for charge $-1$ in Supplementary Figures 10 and 11 (see the analytical solutions in Supplementary Information IV). Along the 12 edges of the charge $-1$ octahedron (Supplementary Figure 10a), the evolution shows strong resemblance to the three-band models[22]. It is because only three bands participate in the edge state pumping. As such, all these transitions can be understood via the rotations of eigenstates in the subgroup $SO(3)$, while the fourth band is fully fixed and decoupled. One other important note-there are 12 possible routes rather than 15 (naively from $C(6,2) = 15$), because it is impossible to evolve directly between two orthogonal planes (or between the diagonal points linked by the dashed lines in Supplementary Figures 10a and b), e.g. between $q_{12}^2$ and $q_{34}^2$. We also find the transition can take arbitrary routes on the 8 faces of the charge $-1$ octahedron (see an example in Supplementary Figures 10d and 11). Supplementary Figure 12 shows the evolution of 2D extended band degeneracies, which help us to understand the pumping of edge states accordingly, e.g. the double-quadratic or triple-linear degeneracies at $k_r = 0$ predict the emergence of topological edge states[22].

**Observation of charges $\pm q_{1234}$ in a transmission line network**

In order to realize and characterize charges $\pm q_{1234}$, we designed a transmission line network[22,27,28] (see the sample photo in Supplementary Figure 13) consisting of 11 unit-cells. There are four meta-atoms A, B, C and D in one unit-cell. The real-space Hamiltonian reads (see details in Supplementary Information III),

$$\mathcal{H} = \sum_n \left( \sum_{\substack{X=A,B,C,D \\ Y=A,B,C,D}} s_{XY} c_{X,n}^\dagger c_{Y,n} + \sum_{\substack{X=A,B,C,D \\ Y=A,B,C,D}} v_{XY} c_{X,n}^\dagger c_{Y,n+1} + \mathrm{h.c.} \right)$$

where $c_{X,n}^\dagger$ and $c_{X,n}$ are creation and annihilation operators on the sub-lattice '$X/Y$' and site '$n$', respectively. In order to realize an explicitly real Hamiltonian in momentum space, we introduce imaginary hoppings[22]. More details on the experimental realization are provided in Supplementary Information V and ref[22].

The left two panels (BulkS) in Fig. 4a show respectively the numerically calculated and experimentally measured energy bands. We plot the corresponding eigenstate trajectories of the four bands in Figs. 4c and d. In the experimental model, at each $k$ we can see one rotation plane is spanned by the eigenvectors of the first and second bands and the other one by those of the third and fourth bands. We can expect that there are two topological edge states in total and one locates in the first bandgap (sandwiched by the first and second bands) while the other one in the third bandgap (formed by the third and fourth bands). The rightmost panels (EdgeS) of Fig. 4a confirm our expectation. The detailed field distribution is provided in Supplementary Figure 14b. The distribution of edge states can also be directly inferred from the 2D extended energy bands as shown in Fig. 4b, where there is one linear Dirac cone between the first/third and second/fourth bands.

**Domain-wall states between charges $+q_{1234}$ and $-q_{1234}$**

If two samples with different non-Abelian topological charges meet at a domain-wall[22], some domain-wall states (DMS) will emerge and their existence can be predicted by defining a "domain-wall charge" $\Delta Q = Q_L/Q_R$. Here $Q_L$ and $Q_R$ are the non-Abelian topological charges of the left and right samples, respectively. The quotient charge $\Delta Q$ is also an element of the non-Abelian group and governs the properties (including both location and number) of the DWS. We note that the appearance of the domain-wall charge -1 in the three-band system can only be well defined by assuming a joint $k$-space basepoint between the left and right samples[22]. Otherwise one cannot distinguish two non-Abelian topological charges (e.g. $+i$ and $-i$) in the same conjugacy class of a three-band system, and thus the domain-wall charge becomes ill-defined. In the four-band system, however, there exists a basepoint-free domain-wall charge taking value of $-1$ between charges $+q_{1234}$ and $-q_{1234}$. It is because they belong to two different conjugacy classes.

In the experiment, we construct a domain-wall (blue spheres in Fig. 4e) between charges $\pm q_{1234}$ as shown in Fig. 4e, where we flip the directions of imaginary hoppings between meta-atoms C and D as denoted by the blue arrows to realize the charge $+q_{1234}$ at the righthand side of the domain wall. Figure 4f shows the domain-wall states between them, where the left inset is the simulated energy levels and the right two insets indicate the measured spectra on the domain-wall for two different excitation/probe locations accordingly. These results indicate that there are two nearly degenerate topological domain-wall states in the third bandgap. This is the same as the hard boundary edge states of charge $-1$ and thus confirms our prediction. The detailed field distribution is provided in Supplementary Figure 15.

**Observation of charges $\pm q_{14}$ in a transmission line network**

In addition, we also experimentally studied charges $\pm q_{14}$, which is also interesting in the four-band models as it exhibits three edge states in the three bandgaps. As shown in Fig. 5, from the bulk bands (Fig. 5a-BulkS), edge state distributions (Fig. 5a-EdgeS) and eigenstate trajectories (Fig. 5c-d), the numerical calculations correctly predict the experimental results. Different from Fig. 1b ($\pm q_{14}$) of the flat band model, the 2D extended energy bands in Fig. 5b are bridged by three linear Dirac cones. As mentioned above, each implies one edge state (per edge), as verified in Fig. 5a-EdgeS. For charge $\pm q_{14}$, there is no complete bandgap in the 2D extended bands. It can be regarded as the generalization of charge $\pm j$ in three-band models[9,22].

**Discussions and conclusion**

Other general configurations of charges $\pm q_{1234}$ are shown in Supplementary Figures 16 and 17, corresponding to the factorizations of $-q_{13}q_{24}$ and $-q_{14}q_{23}$, respectively. As mentioned above, the ring degeneracy formed by the second and third bands in Fig. 1c ($-q_{13}q_{24}$) splits into two Dirac cones as shown in Supplementary Figure 16c, which further imply two edge states (per edge) in Supplementary Figure 16d. The general model of charge $-1$ in Supplementary Figure 18 shows one triple linear degeneracy, being similar to what we have seen in the three-band models, as such the edge state distributions[22].

Our exhaustive study of all non-Abelian topological charges of the PT symmetric four-band Hamiltonians will constructively stimulate the related research on 2D systems, e.g. twisted bi-layer graphene[10,29,30]. The PT symmetric systems also may contribute to exotic fragile topological states[31] and even topological effective gravitational theory[32]. The studies can be easily transferred to other artificial platforms including optical lattices[33], photonics[34-36] and phononics[37].

## Table and Figures

Table 1. Category of non-Abelian topological charges in four-band models. The three categories could be further decomposed into 10 conjugacy classes forming the generalized quaternion group $Q_{16}$. For the four-band system separated by three bandgaps, if we label each band with Zak phases of 0 or $\pi$, there are $2^3 = 8$ possibilities, corresponding to the 8 different eigenvalue sets. There two classes going beyond the Zak phase description[9].

| $Q_{16}$: Clifford-basis label | $Q_{16}$: Band-index label | Eigenvalues: $(\lambda_1, \lambda_2, \lambda_3, \lambda_4)$ |
|---|---|---|
| $\{+1\}, \{-1\}$ | $\{+1\}, \{-1\}$ | $(1,1,1,1)$ |
| $\{\pm e_1\}$<br>$\{\pm e_2\}$<br>$\{\pm e_3\}$<br>$\{\pm e_1 e_2\}$<br>$\{\pm e_1 e_3\}$<br>$\{\pm e_2 e_3\}$ | $\{\pm q_{12}\}$<br>$\{\pm q_{13}\}$<br>$\{\pm q_{14}\}$<br>$\{\pm q_{23}\}$<br>$\{\pm q_{24}\}$<br>$\{\pm q_{34}\}$ | $(-1,-1,1,1)$<br>$(-1,1,-1,1)$<br>$(-1,1,1,-1)$<br>$(1,-1,-1,1)$<br>$(1,-1,1,-1)$<br>$(1,1,-1,-1)$ |
| $\{+e_1 e_2 e_3\}, \{-e_1 e_2 e_3\}$ | $\{+q_{1234}\}, \{-q_{1234}\}$ | $(-1,-1,-1,-1)$ |

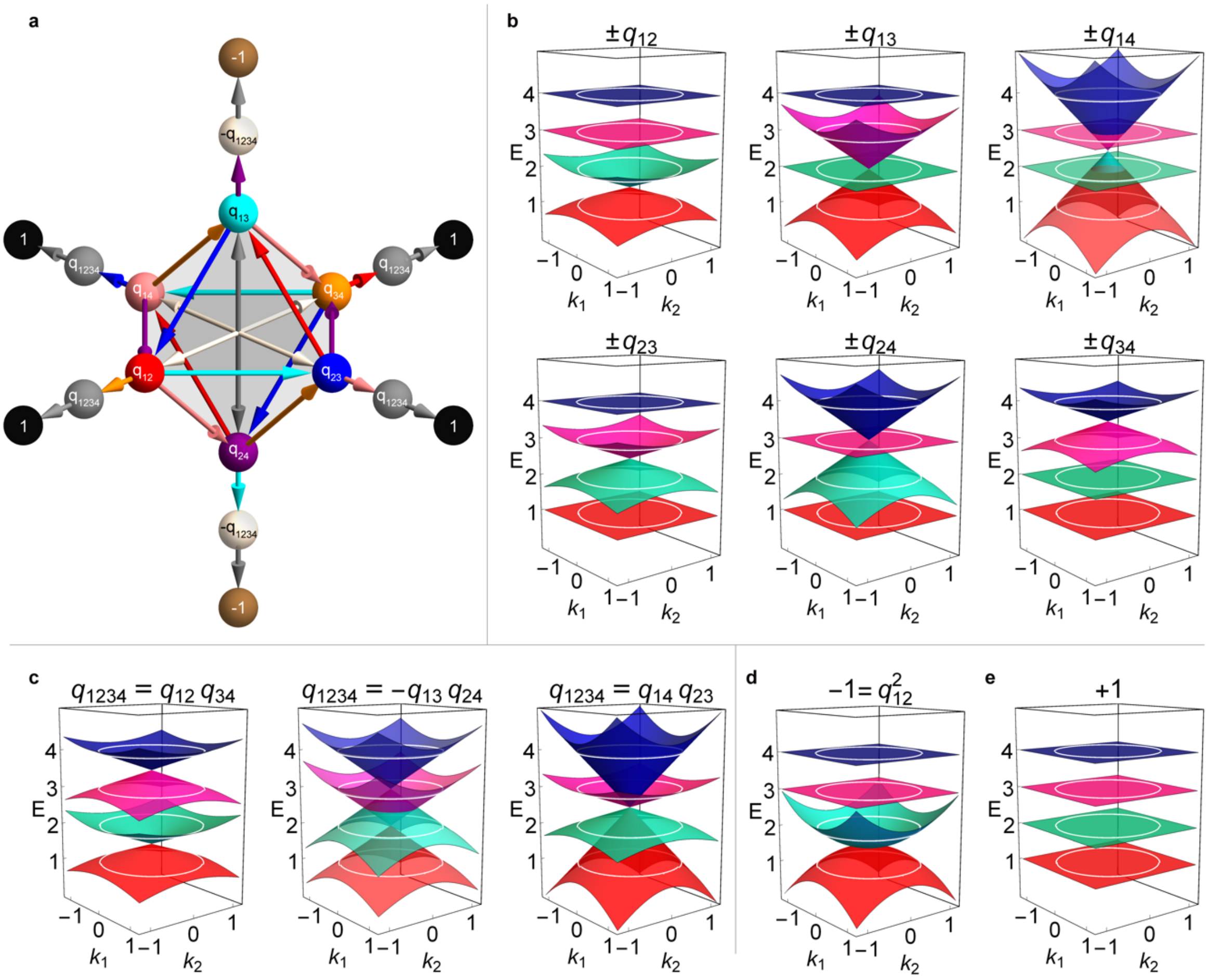

**Figure 1. Non-Abelian topological charges in four-band models. a,** Elements of the $Q_{16}$ group indicated by coloured spheres sitting on an outstretched regular octahedron and their mutual multiplications represented by the corresponding coloured arrows. For example, a red arrow $q_{12}$ brings a blue sphere $q_{23}$ to a cyan sphere $q_{13}$ indicating that $q_{23}q_{12} = q_{13}$. Full multiplication tables are provided in Supplementary Tables 1 and 2. **b-e,** The extended two-dimensional bands corresponding to three different categories of the non-Abelian topological charges: $\pm q_{mn}$, $q_{1234}$ and $\pm 1$, respectively. White circles indicate the corresponding 1D bands. The charge $q_{1234}$ can be decomposed into three ways in (c), which are all topologically equivalent. For charge $-1$ we take $-1 = q_{12}^2$ as an example, and the other cases can be simply obtained by changing the linear Dirac cone degeneracies in panel (b) to be quadratic ones without any position shifting.

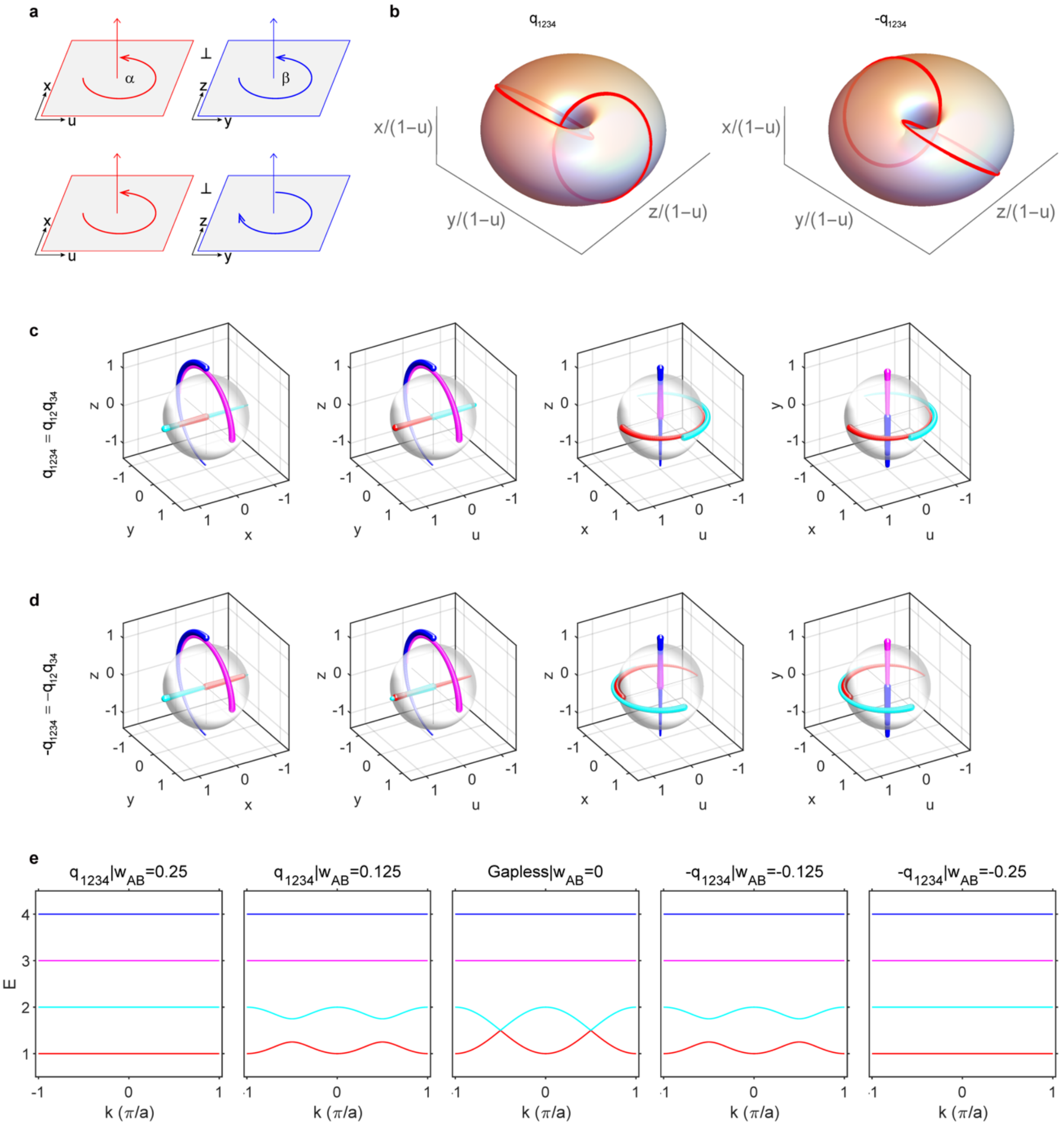

**Figure 2. Non-Abelian topological charges $\pm q_{1234}$ specific to four-band models. a,** Rotations in four-dimension. For each rotation $R$, there is at least one pair of two orthogonal rotation-invariant planes, e.g. $A = oux$ and $B = oyz$, which span the four-dimensional space. For any $\vec{a} \in A$ and $\vec{b} \in B$ we have $\vec{a} \perp \vec{b}$, $R\vec{a} \in A$ and $R\vec{b} \in B$. We define the angle between $\vec{a}$ and $R\vec{a}$ ($\vec{b}$ and $R\vec{b}$) in the plane $A(B)$ as $\alpha(\beta)$. **b,** Four-dimensional Clifford tori $(u, x, y, z)$ are stereographically projected into $\mathbb{R}^3$ as the conventional tori $\left(\frac{x}{1-u}, \frac{y}{1-u}, \frac{z}{1-u}\right)$. The two linked circles represent the trajectory of one eigenstate (with applying all the four-dimensional $D_2$ rotations), the other three eigenstate trajectories overlap with this one. **c and d,** Orthographically projecting the four eigenstate trajectories (shown in different colours) onto four 3D solid spheres, where one component is hidden by projection in each sphere. **e,** The topological phase transition between charges $\pm q_{1234}$, other parameters are listed in Supplementary Table 4.

**a**

| $Q_{16}$ | $\pm q_{12}$ | $\pm q_{13}$ | $\pm q_{14}$ | $\pm q_{23}$ | $\pm q_{24}$ | $\pm q_{34}$ | $+1$ | $-1$ | $+q_{1234}$ | $-q_{1234}$ |
|---|---|---|---|---|---|---|---|---|---|---|
| 4th band | 0 | 0 | $\pm\pi$ | 0 | $\pm\pi$ | $\pm\pi$ | 0 | $2\pi$ | Two $+\pi$ | Two $+\pi$ |
| 3rd band | 0 | $\pm\pi$ | 0 | $\pm\pi$ | 0 | $\pm\pi$ | | | Two $\pm\pi$ | Two $\mp\pi$ |
| 2nd band | $\pm\pi$ | 0 | 0 | $\pm\pi$ | $\pm\pi$ | 0 | | | | |
| 1st band | $\pm\pi$ | $\pm\pi$ | $\pm\pi$ | 0 | 0 | 0 | | | | |

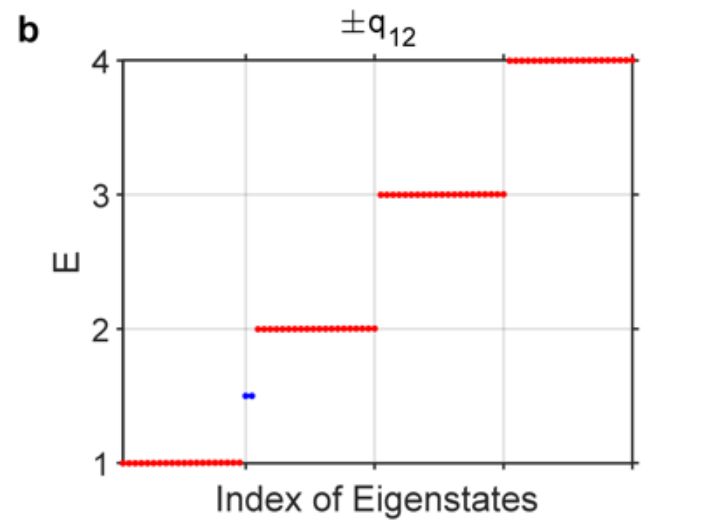

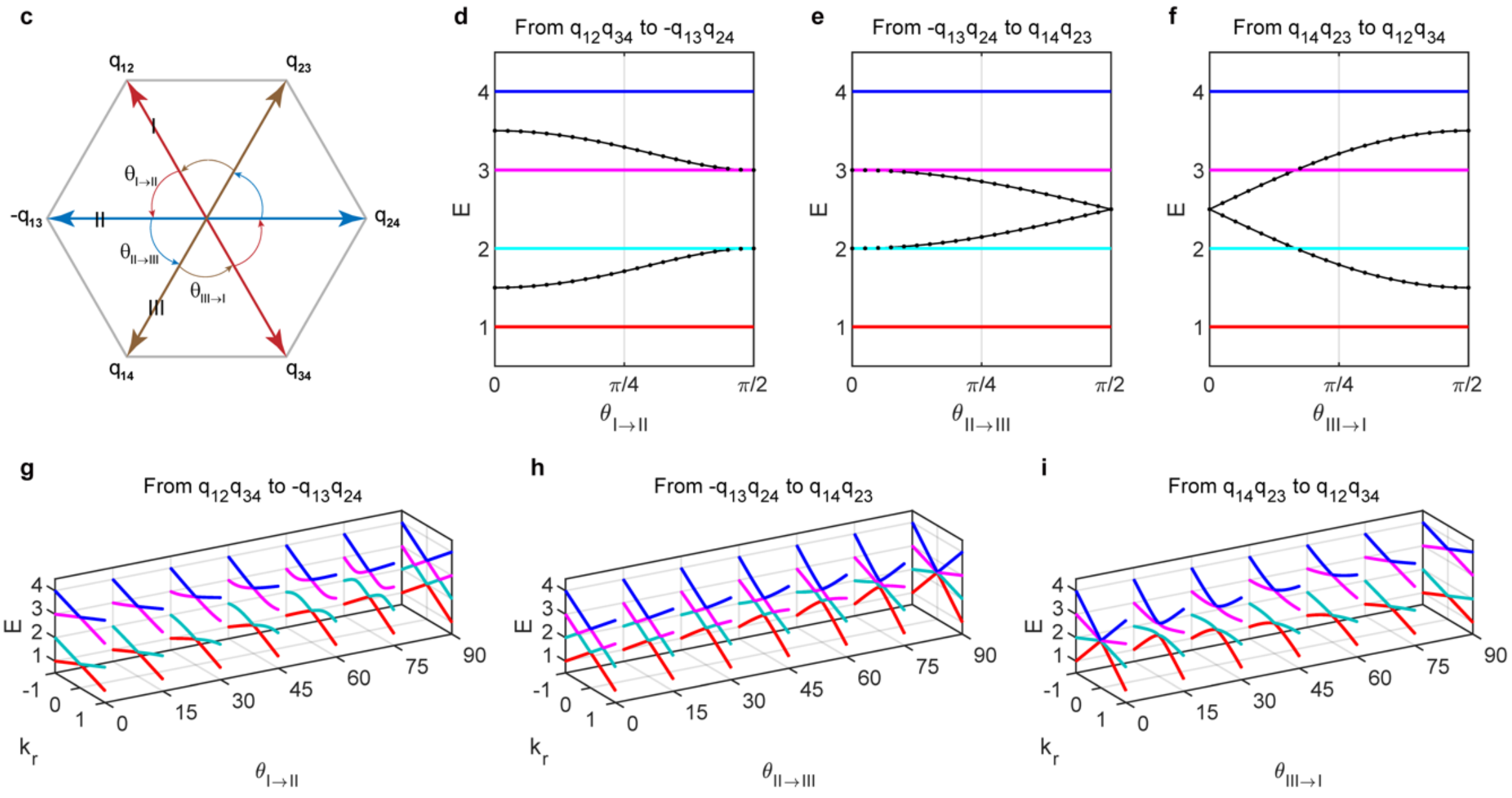

**Figure 3. Zak phases and evolution of edge states. a,** Zak phases for individual bands summarized for each non-Abelian topological charge. For charges $\pm q_{1234}$, two eigenvectors rotate $\pi$, while the other two rotate $+\pi$ or $-\pi$ depending on the factorizations (due to the handedness of subspace). **b,** Edge states of charges $\pm q_{12}$ occur at the bandgap sandwiched by the first and second bands. **c,** Schematic view of charge $q_{1234}$ factorizations and their mutual continuous transitions. The double-headed arrows indicate that the paired two factors commute, i.e. $q_{12}q_{34} = q_{34}q_{12} = q_{1234}$. The directional arcs define the continuous transitions parametrized by $\theta_{m\to n}$ with $m$ and $n$ taking values of I, II and III, corresponding to the factorizations of $q_{12}q_{34}$, $-q_{13}q_{24}$ and $q_{14}q_{23}$, respectively. **d-f,** Evolution of edge states with varying the parameters $\theta_{m\to n}$, where lines/dots indicate the numerical/analytical results. **g-i,** Evolution of the extended two-dimensional bands corresponding to (d-f), respectively. Note that we only plot the radial cuts $E(k_r)$ as the 2D bands for ideal flat-band models are isotropic in the $(k_1, k_2)$ plane.

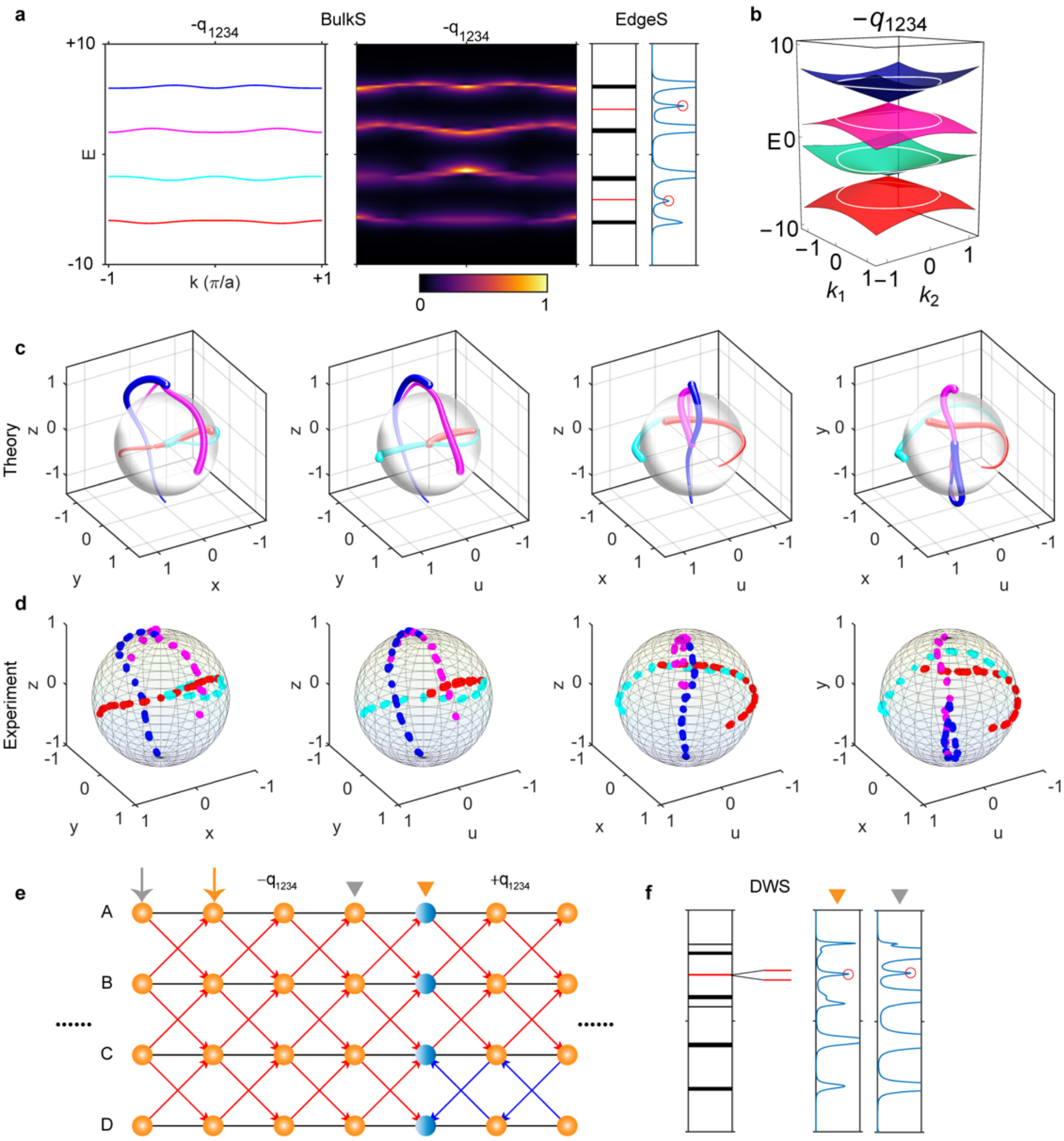

**Figure 4. Experimental observation of the charge $-q_{1234}$ and edge/domain-wall states. a,** The left panels show the numerically calculated and experimentally measured energy bands of bulk states (BulkS). The right panels indicate the energy spectra probed at the hard boundary, where the red lines and the peaks marked by red circles represent the simulation and experiment results of edge states (EdgeS), respectively. **b,** The extended energy bands on a 2D plane, there is one linear Dirac cone between the first/third and second/fourth bands. White circles indicate the 1D energy bands. **c/d,** The calculated/measured orthographic projections of eigenstate trajectories. The colours of trajectories correspond to different bands in (a), respectively. The direction of linewidth decreasing indicates $k$ running from $-\pi$ to $\pi$. **e,** The construction of domain-wall (marked by blue spheres) between charges $-q_{1234}$ and $q_{1234}$. The gray and yellow arrows/triangles denote two different excitation/probe positions. **f,** The calculated and measured energy spectra for the two different pairs of excitation and probe positions,

respectively. The two domain-wall states are nearly degenerate. Detailed distribution of edge/domain-wall states is shown in Supplementary Figure 14/15.

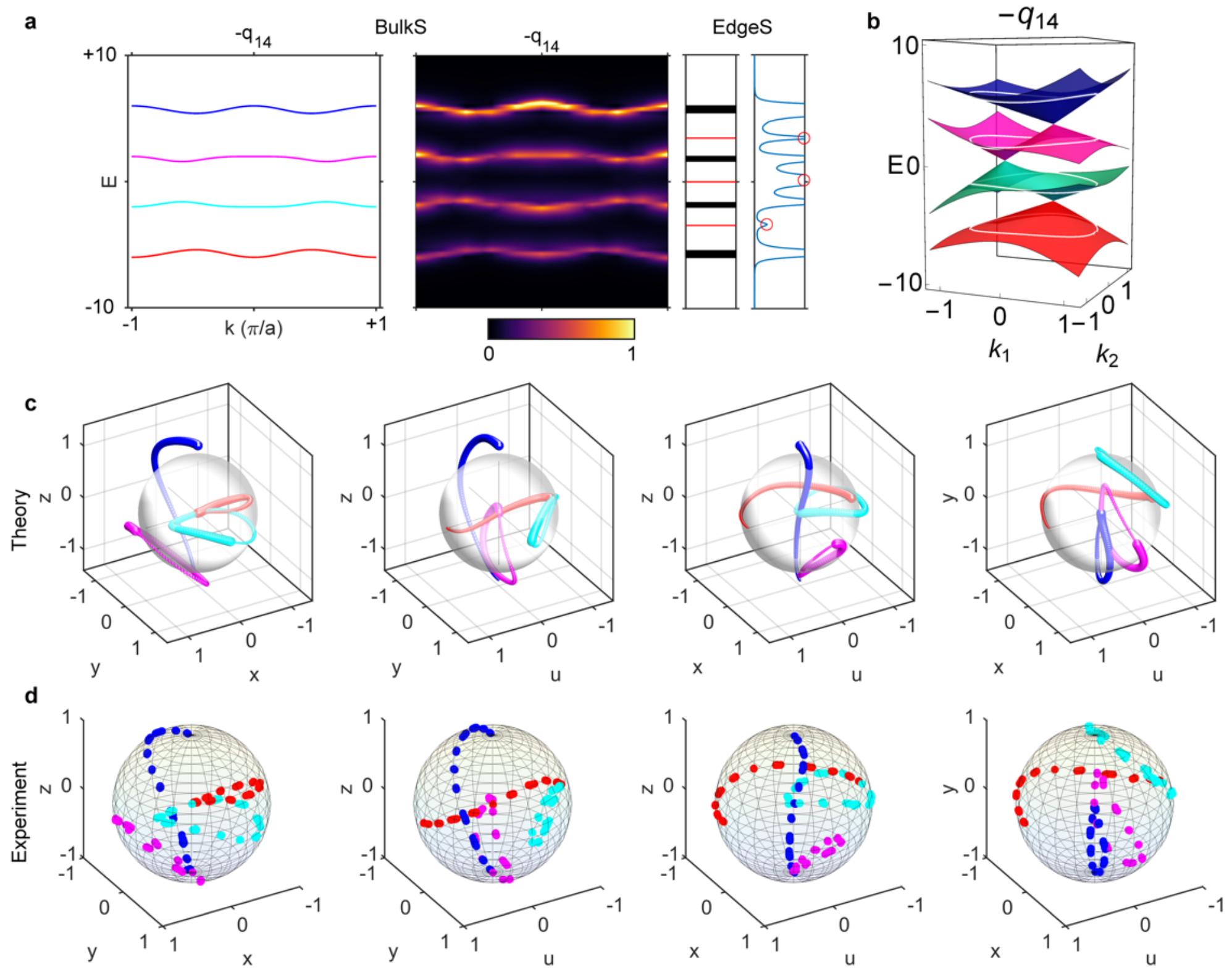

**Figure 5. Experimental observation of the charge $-q_{14}$ and edge states. a,** The left panels show the numerically calculated and experimentally measured energy bands of bulk states (BulkS). The right panels indicate the energy spectra probed at the hard boundary, where the red lines and the peaks marked by red circles represent the simulation and experiment results of edge states (EdgeS), respectively. **b,** The extended energy bands on a 2D plane, there is one linear Dirac cone in each bandgap. White circles indicate the 1D energy bands. **c/d,** The calculated/measured orthographic projections of eigenstate trajectories. The colours of trajectories correspond to different bands in (a), respectively. The direction of linewidth decreasing indicates $k$ running from $-\pi$ to $\pi$.

**Acknowledgements**

This work is supported by the Hong Kong RGC (AoE/P-02/12, 16310420), the Hong Kong Scholars Program (XJ2019007), the KAUST CRG grant (KAUST20SC01) and the Croucher foundation (CAS20SC01).

# Supplementary Information:

## Contents

# 1. Non-Abelian topological charges in four-band models

The calculation[9] of non-Abelian topological charges requires lifting the Berry connection one-form from the Lie algebra $\mathfrak{so}(4)$ to $\mathfrak{spin}(4)$, i.e.,

$$L_{ij} \to t_{ij} = -\frac{1}{4}\left[\Gamma_i, \Gamma_j\right] \in \mathfrak{spin}(4) \tag{S1}$$

where the corresponding coefficients $\beta_a^{ij}$ are kept as,

$$[A(k)]_a = \sum_{i<j} \beta_a^{ij}(k) L_{ij} \xrightarrow{lift} [\bar{A}(k)]_a = \sum_{i<j} \beta_a^{ij}(k) t_{ij} \tag{S2}$$

We set the basis of the Lie algebra $\mathfrak{so}(4)$ as $\left(L_{ij}\right)_{a,b=1:4} = -\delta_{ia}\delta_{jb} + \delta_{ib}\delta_{ja}$, the explicit forms are,

$$\mathrm{L}_{12} = \begin{pmatrix} 0 & -1 & 0 & 0 \\ 1 & 0 & 0 & 0 \\ 0 & 0 & 0 & 0 \\ 0 & 0 & 0 & 0 \end{pmatrix}, \mathrm{L}_{13} = \begin{pmatrix} 0 & 0 & -1 & 0 \\ 0 & 0 & 0 & 0 \\ 1 & 0 & 0 & 0 \\ 0 & 0 & 0 & 0 \end{pmatrix}, \mathrm{L}_{14} = \begin{pmatrix} 0 & 0 & 0 & -1 \\ 0 & 0 & 0 & 0 \\ 0 & 0 & 0 & 0 \\ 1 & 0 & 0 & 0 \end{pmatrix};$$

$$\mathrm{L}_{23} = \begin{pmatrix} 0 & 0 & 0 & 0 \\ 0 & 0 & -1 & 0 \\ 0 & 1 & 0 & 0 \\ 0 & 0 & 0 & 0 \end{pmatrix}, \mathrm{L}_{24} = \begin{pmatrix} 0 & 0 & 0 & 0 \\ 0 & 0 & 0 & -1 \\ 0 & 0 & 0 & 0 \\ 0 & 1 & 0 & 0 \end{pmatrix}, \mathrm{L}_{34} = \begin{pmatrix} 0 & 0 & 0 & 0 \\ 0 & 0 & 0 & 0 \\ 0 & 0 & 0 & -1 \\ 0 & 0 & 1 & 0 \end{pmatrix} \tag{S3}$$

We use the following $\Gamma$ matrices,

$$\Gamma_1 = \begin{pmatrix} 0 & 0 & 0 & i \\ 0 & 0 & i & 0 \\ 0 & -i & 0 & 0 \\ -i & 0 & 0 & 0 \end{pmatrix}, \Gamma_2 = \begin{pmatrix} 0 & 0 & 0 & 1 \\ 0 & 0 & -1 & 0 \\ 0 & -1 & 0 & 0 \\ 1 & 0 & 0 & 0 \end{pmatrix}$$

$$\Gamma_3 = \begin{pmatrix} 0 & 0 & i & 0 \\ 0 & 0 & 0 & -i \\ -i & 0 & 0 & 0 \\ 0 & i & 0 & 0 \end{pmatrix}, \Gamma_4 = \begin{pmatrix} 0 & 0 & 1 & 0 \\ 0 & 0 & 0 & 1 \\ 1 & 0 & 0 & 0 \\ 0 & 1 & 0 & 0 \end{pmatrix} \tag{S4}$$

They are obtained from,

$$\Gamma_1 = -\sigma_2 \otimes \sigma_1, \Gamma_2 = -\sigma_2 \otimes \sigma_2, \Gamma_3 = -\sigma_2 \otimes \sigma_3, \Gamma_4 = \sigma_1 \otimes \sigma_0 \tag{S5}$$

where $\sigma_i (i = 1,2,3)$ are Pauli matrices and $\sigma_0$ is the $2 \times 2$ identity matrix. They satisfy the anticommutation relations $\left\{\Gamma_i, \Gamma_j\right\} = 2\delta_{ij}$.

The basis of Clifford algebra $C\ell_{0,3}$ that generates the group $Q_{16}$ is obtained by,

$$e_{j-1} \equiv \exp\left(-\frac{\pi}{4}\left[\Gamma_1, \Gamma_j\right]\right) = \frac{1}{2}\left[\Gamma_j, \Gamma_1\right] \tag{S6}$$

for $2 \le j \le 4$. Their explicit forms are,

$$e_1 = \frac{1}{2}[\Gamma_2, \Gamma_1] = \begin{pmatrix} -i & 0 & 0 & 0 \\ 0 & i & 0 & 0 \\ 0 & 0 & -i & 0 \\ 0 & 0 & 0 & i \end{pmatrix}$$

$$e_2 = \frac{1}{2}[\Gamma_3, \Gamma_1] = \begin{pmatrix} 0 & 1 & 0 & 0 \\ -1 & 0 & 0 & 0 \\ 0 & 0 & 0 & 1 \\ 0 & 0 & -1 & 0 \end{pmatrix}$$

$$e_3 = \frac{1}{2}[\Gamma_4, \Gamma_1] = \begin{pmatrix} 0 & -i & 0 & 0 \\ -i & 0 & 0 & 0 \\ 0 & 0 & 0 & i \\ 0 & 0 & i & 0 \end{pmatrix} \tag{S7}$$

with $e_i^2 = -I_{4\times4}$ and $e_i e_j = -e_j e_i$ (or we combine them two together as $\{e_i, e_j\} = -2\delta_{ij}$).

From the basis $\{e_1, e_2, e_3\}$, we have,

$$Q_{16} = \bigcup_{n_i \in \{0,1\}} \{\pm e_1^{n_1} e_2^{n_2} e_3^{n_3}\}$$

$$= \{\pm e_1, \pm e_2, \pm e_3, \pm e_1 e_2, \pm e_1 e_3, \pm e_2 e_3\} \cup \{+1, -1, +e_1 e_2 e_3, -e_1 e_2 e_3\} \tag{S8}$$

where,

$$e_{12} \equiv e_1 e_2 = \begin{pmatrix} 0 & -i & 0 & 0 \\ -i & 0 & 0 & 0 \\ 0 & 0 & 0 & -i \\ 0 & 0 & -i & 0 \end{pmatrix}, e_{13} \equiv e_1 e_3 = \begin{pmatrix} 0 & -1 & 0 & 0 \\ 1 & 0 & 0 & 0 \\ 0 & 0 & 0 & 1 \\ 0 & 0 & -1 & 0 \end{pmatrix}$$

$$e_{23} \equiv e_2 e_3 = \begin{pmatrix} -i & 0 & 0 & 0 \\ 0 & i & 0 & 0 \\ 0 & 0 & i & 0 \\ 0 & 0 & 0 & -i \end{pmatrix}, e_{123} \equiv e_1 e_2 e_3 = \begin{pmatrix} -1 & 0 & 0 & 0 \\ 0 & -1 & 0 & 0 \\ 0 & 0 & 1 & 0 \\ 0 & 0 & 0 & 1 \end{pmatrix} \tag{S9}$$

We rename each $Q_{16}$ group element as (to be more convenient in physics analysis; also see Table 1 in the main text),

$$e_1 \to q_{12}, e_2 \to q_{13}, e_3 \to q_{14}, e_{12} \to q_{23}, e_{13} \to q_{24}, e_{23} \to q_{34}, e_{123} \to q_{1234} \tag{S10}$$

Finally, we have the multiplication tables as shown in Tables S1 and S2.

Table S1. Multiplication table of $Q_{16}$ group labelled with Clifford algebra basis.

$Q_{16}$ multiplication table

| | +1 | $e_1$ | $e_2$ | $e_3$ | $e_{12}$ | $e_{13}$ | $e_{23}$ | $e_{123}$ | -1 | $-e_1$ | $-e_2$ | $-e_3$ | $-e_{12}$ | $-e_{13}$ | $-e_{23}$ | $-e_{123}$ |
|---|---|---|---|---|---|---|---|---|---|---|---|---|---|---|---|---|
| +1 | +1 | $e_1$ | $e_2$ | $e_3$ | $e_{12}$ | $e_{13}$ | $e_{23}$ | $e_{123}$ | -1 | $-e_1$ | $-e_2$ | $-e_3$ | $-e_{12}$ | $-e_{13}$ | $-e_{23}$ | $-e_{123}$ |
| $e_1$ | $e_1$ | -1 | $e_{12}$ | $e_{13}$ | $-e_2$ | $-e_3$ | $e_{123}$ | $-e_{23}$ | $-e_1$ | +1 | $-e_{12}$ | $-e_{13}$ | $e_2$ | $e_3$ | $-e_{123}$ | $e_{23}$ |
| $e_2$ | $e_2$ | $-e_{12}$ | -1 | $e_{23}$ | $e_1$ | $-e_{123}$ | $-e_3$ | $e_{13}$ | $-e_2$ | $e_{12}$ | +1 | $-e_{23}$ | $-e_1$ | $e_{123}$ | $e_3$ | $-e_{13}$ |
| $e_3$ | $e_3$ | $-e_{13}$ | $-e_{23}$ | -1 | $e_{123}$ | $e_1$ | $e_2$ | $-e_{12}$ | $-e_3$ | $e_{13}$ | $e_{23}$ | +1 | $-e_{123}$ | $-e_1$ | $-e_2$ | $e_{12}$ |
| $e_{12}$ | $e_{12}$ | $e_2$ | $-e_1$ | $e_{123}$ | -1 | $e_{23}$ | $-e_{13}$ | $-e_3$ | $-e_{12}$ | $-e_2$ | $e_1$ | $-e_{123}$ | +1 | $-e_{23}$ | $e_{13}$ | $e_3$ |
| $e_{13}$ | $e_{13}$ | $e_3$ | $-e_{123}$ | $-e_1$ | $-e_{23}$ | -1 | $e_{12}$ | $e_2$ | $-e_{13}$ | $-e_3$ | $e_{123}$ | $e_1$ | $e_{23}$ | +1 | $-e_{12}$ | $-e_2$ |
| $e_{23}$ | $e_{23}$ | $e_{123}$ | $e_3$ | $-e_2$ | $e_{13}$ | $-e_{12}$ | -1 | $-e_1$ | $-e_{23}$ | $-e_{123}$ | $-e_3$ | $e_2$ | $-e_{13}$ | $e_{12}$ | +1 | $e_1$ |
| $e_{123}$ | $e_{123}$ | $-e_{23}$ | $e_{13}$ | $-e_{12}$ | $-e_3$ | $e_2$ | $-e_1$ | +1 | $-e_{123}$ | $e_{23}$ | $-e_{13}$ | $e_{12}$ | $e_3$ | $-e_2$ | $e_1$ | -1 |
| -1 | -1 | $-e_1$ | $-e_2$ | $-e_3$ | $-e_{12}$ | $-e_{13}$ | $-e_{23}$ | $-e_{123}$ | +1 | $e_1$ | $e_2$ | $e_3$ | $e_{12}$ | $e_{13}$ | $e_{23}$ | $e_{123}$ |
| $-e_1$ | $-e_1$ | +1 | $-e_{12}$ | $-e_{13}$ | $e_2$ | $e_3$ | $-e_{123}$ | $e_{23}$ | $e_1$ | -1 | $e_{12}$ | $e_{13}$ | $-e_2$ | $-e_3$ | $e_{123}$ | $-e_{23}$ |
| $-e_2$ | $-e_2$ | $e_{12}$ | +1 | $-e_{23}$ | $-e_1$ | $e_{123}$ | $e_3$ | $-e_{13}$ | $e_2$ | $-e_{12}$ | -1 | $e_{23}$ | $e_1$ | $-e_{123}$ | $-e_3$ | $e_{13}$ |
| $-e_3$ | $-e_3$ | $e_{13}$ | $e_{23}$ | +1 | $-e_{123}$ | $-e_1$ | $-e_2$ | $e_{12}$ | $e_3$ | $-e_{13}$ | $-e_{23}$ | -1 | $e_{123}$ | $e_1$ | $e_2$ | $-e_{12}$ |
| $-e_{12}$ | $-e_{12}$ | $-e_2$ | $e_1$ | $-e_{123}$ | +1 | $-e_{23}$ | $e_{13}$ | $e_3$ | $e_{12}$ | $e_2$ | $-e_1$ | $e_{123}$ | -1 | $e_{23}$ | $-e_{13}$ | $-e_3$ |
| $-e_{13}$ | $-e_{13}$ | $-e_3$ | $e_{123}$ | $e_1$ | $e_{23}$ | +1 | $-e_{12}$ | $-e_2$ | $e_{13}$ | $e_3$ | $-e_{123}$ | $-e_1$ | $-e_{23}$ | -1 | $e_{12}$ | $e_2$ |
| $-e_{23}$ | $-e_{23}$ | $-e_{123}$ | $-e_3$ | $e_2$ | $-e_{13}$ | $e_{12}$ | +1 | $e_1$ | $e_{23}$ | $e_{123}$ | $e_3$ | $-e_2$ | $e_{13}$ | $-e_{12}$ | -1 | $-e_1$ |
| $-e_{123}$ | $-e_{123}$ | $e_{23}$ | $-e_{13}$ | $e_{12}$ | $e_3$ | $-e_2$ | $e_1$ | -1 | $e_{123}$ | $-e_{23}$ | $e_{13}$ | $-e_{12}$ | $-e_3$ | $e_2$ | $-e_1$ | +1 |

Table S2. Multiplication table of $Q_{16}$ group labelled with band-index.

$Q_{16}$ multiplication table

| | +1 | $q_{12}$ | $q_{13}$ | $q_{14}$ | $q_{23}$ | $q_{24}$ | $q_{34}$ | $q_{1234}$ | -1 | $-q_{12}$ | $-q_{13}$ | $-q_{14}$ | $-q_{23}$ | $-q_{24}$ | $-q_{34}$ | $-q_{1234}$ |
|---|---|---|---|---|---|---|---|---|---|---|---|---|---|---|---|---|
| +1 | +1 | $q_{12}$ | $q_{13}$ | $q_{14}$ | $q_{23}$ | $q_{24}$ | $q_{34}$ | $q_{1234}$ | -1 | $-q_{12}$ | $-q_{13}$ | $-q_{14}$ | $-q_{23}$ | $-q_{24}$ | $-q_{34}$ | $-q_{1234}$ |
| $q_{12}$ | $q_{12}$ | -1 | $q_{23}$ | $q_{24}$ | $-q_{13}$ | $-q_{14}$ | $q_{1234}$ | $-q_{34}$ | $-q_{12}$ | +1 | $-q_{23}$ | $-q_{24}$ | $q_{13}$ | $q_{14}$ | $-q_{1234}$ | $q_{34}$ |
| $q_{13}$ | $q_{13}$ | $-q_{23}$ | -1 | $q_{34}$ | $q_{12}$ | $-q_{1234}$ | $-q_{14}$ | $q_{24}$ | $-q_{13}$ | $q_{23}$ | +1 | $-q_{34}$ | $-q_{12}$ | $q_{1234}$ | $q_{14}$ | $-q_{24}$ |
| $q_{14}$ | $q_{14}$ | $-q_{24}$ | $-q_{34}$ | -1 | $q_{1234}$ | $q_{12}$ | $q_{13}$ | $-q_{23}$ | $-q_{14}$ | $q_{24}$ | $q_{34}$ | +1 | $-q_{1234}$ | $-q_{12}$ | $-q_{13}$ | $q_{23}$ |
| $q_{23}$ | $q_{23}$ | $q_{13}$ | $-q_{12}$ | $q_{1234}$ | -1 | $q_{34}$ | $-q_{24}$ | $-q_{14}$ | $-q_{23}$ | $-q_{13}$ | $q_{12}$ | $-q_{1234}$ | +1 | $-q_{34}$ | $q_{24}$ | $q_{14}$ |
| $q_{24}$ | $q_{24}$ | $q_{14}$ | $-q_{1234}$ | $-q_{12}$ | $-q_{34}$ | -1 | $q_{23}$ | $q_{13}$ | $-q_{24}$ | $-q_{14}$ | $q_{1234}$ | $q_{12}$ | $q_{34}$ | +1 | $-q_{23}$ | $-q_{13}$ |
| $q_{34}$ | $q_{34}$ | $q_{1234}$ | $q_{14}$ | $-q_{13}$ | $q_{24}$ | $-q_{23}$ | -1 | $-q_{12}$ | $-q_{34}$ | $-q_{1234}$ | $-q_{14}$ | $q_{13}$ | $-q_{24}$ | $q_{23}$ | +1 | $q_{12}$ |
| $q_{1234}$ | $q_{1234}$ | $-q_{34}$ | $q_{24}$ | $-q_{23}$ | $-q_{14}$ | $q_{13}$ | $-q_{12}$ | +1 | $-q_{1234}$ | $q_{34}$ | $-q_{24}$ | $q_{23}$ | $q_{14}$ | $-q_{13}$ | $q_{12}$ | -1 |
| -1 | -1 | $-q_{12}$ | $-q_{13}$ | $-q_{14}$ | $-q_{23}$ | $-q_{24}$ | $-q_{34}$ | $-q_{1234}$ | +1 | $q_{12}$ | $q_{13}$ | $q_{14}$ | $q_{23}$ | $q_{24}$ | $q_{34}$ | $q_{1234}$ |
| $-q_{12}$ | $-q_{12}$ | +1 | $-q_{23}$ | $-q_{24}$ | $q_{13}$ | $q_{14}$ | $-q_{1234}$ | $q_{34}$ | $q_{12}$ | -1 | $q_{23}$ | $q_{24}$ | $-q_{13}$ | $-q_{14}$ | $q_{1234}$ | $-q_{34}$ |
| $-q_{13}$ | $-q_{13}$ | $q_{23}$ | +1 | $-q_{34}$ | $-q_{12}$ | $q_{1234}$ | $q_{14}$ | $-q_{24}$ | $q_{13}$ | $-q_{23}$ | -1 | $q_{34}$ | $q_{12}$ | $-q_{1234}$ | $-q_{14}$ | $q_{24}$ |
| $-q_{14}$ | $-q_{14}$ | $q_{24}$ | $q_{34}$ | +1 | $-q_{1234}$ | $-q_{12}$ | $-q_{13}$ | $q_{23}$ | $q_{14}$ | $-q_{24}$ | $-q_{34}$ | -1 | $q_{1234}$ | $q_{12}$ | $q_{13}$ | $-q_{23}$ |
| $-q_{23}$ | $-q_{23}$ | $-q_{13}$ | $q_{12}$ | $-q_{1234}$ | +1 | $-q_{34}$ | $q_{24}$ | $q_{14}$ | $q_{23}$ | $q_{13}$ | $-q_{12}$ | $q_{1234}$ | -1 | $q_{34}$ | $-q_{24}$ | $-q_{14}$ |
| $-q_{24}$ | $-q_{24}$ | $-q_{14}$ | $q_{1234}$ | $q_{12}$ | $q_{34}$ | +1 | $-q_{23}$ | $-q_{13}$ | $q_{24}$ | $q_{14}$ | $-q_{1234}$ | $-q_{12}$ | $-q_{34}$ | -1 | $q_{23}$ | $q_{13}$ |
| $-q_{34}$ | $-q_{34}$ | $-q_{1234}$ | $-q_{14}$ | $q_{13}$ | $-q_{24}$ | $q_{23}$ | +1 | $q_{12}$ | $q_{34}$ | $q_{1234}$ | $q_{14}$ | $-q_{13}$ | $q_{24}$ | $-q_{23}$ | -1 | $-q_{12}$ |
| $-q_{1234}$ | $-q_{1234}$ | $q_{34}$ | $-q_{24}$ | $q_{23}$ | $q_{14}$ | $-q_{13}$ | $q_{12}$ | -1 | $q_{1234}$ | $-q_{34}$ | $q_{24}$ | $-q_{23}$ | $-q_{14}$ | $q_{13}$ | $-q_{12}$ | +1 |

# 2. Rotations in four-dimension

We briefly recall some facts about rotations in the four-dimension[23-25]. For each rotation $R$, there is at least one pair of orthogonal 2-planes (the 2-planes are also dubbed as invariant planes) - $A$ and $B$ which are invariant under the rotation $R$ and span the four-dimensional space, i.e. for any $\vec{a} \in A$ and $\vec{b} \in B$ we have $\vec{a} \perp \vec{b}$, $R\vec{a} \in A$ and $R\vec{b} \in B$. We define the angle between $\vec{a}$ and $R\vec{a}$ ($\vec{b}$ and $R\vec{b}$) in the 2-plane $A$ ($B$) as $\alpha$ ($\beta$). As thus, four-dimensional rotations can be categorized into two types: simple rotations ($\alpha = 0$ and $\beta \neq 0$ or $\alpha \neq 0$ and $\beta = 0$) and double rotations ($\alpha \neq 0$ and $\beta \neq 0$). When the two rotation angles satisfy $|\alpha| = |\beta|$, the rotation $R$ is called isoclinic rotation, where there are infinitely many pairs of orthogonal 2-planes. Assuming that the coordinate set is ordered as $ouxyz$ with $o$ indicating the origin, we consider two 2-planes $A = oux$ and $B = oyz$ and set the rotation angle $\alpha$ ($\beta$) positive from $ou$ to $ox$ ($oy$ to $oz$). Then isoclinic rotations with $\alpha\beta > 0$ are denoted as left-isoclinic; those with $\alpha\beta < 0$ as right-isoclinic. Note that the two cases with $|\alpha| = |\beta| = 0\ or\ \pi$ are the only ones that are simultaneously left- and right-isoclinic. When the Bloch wave-vector runs across the 1D first Brillouin zone ($k = -\pi \to \pi$), the rotation matrix $R(k)$ at each $k$ for each charge can be classified to be single or double rotations. The charges $\pm q_{1234}$ have to be realized via double rotations. In some ideal cases, $+/-\ q_{1234}$ consists of purely left/right-isoclinic rotations, where there are infinitely many pairs of orthogonal 2-planes. All the other charges can be ideally described by simple rotations.

It is well known that rotations can be encoded by quaternion ($\mathbb{H}$) multiplication (discovered by Hamilton in 1843 and frequently used in engineering applications). For example, in three-dimension the rotation $\vec{r}'_{3D} = R_{3D}\vec{r}_{3D}$ can be calculated with $r'_{3D} = q r_{3D} q^{-1}$, where $q \in SU(2)$ being the unit quaternion and $\vec{r}_{3D}$ & $\vec{r}'_{3D} \in \mathbb{R}^3$ can be identified with the pure quaternion (no real part), i.e. $r_{3D}$ & $r'_{3D} \in Im(\mathbb{H})$, respectively. Actually the calculation is made by a surjective homomorphism, $\rho: q \in SU(2) \to R_{3D} \in SO(3)$, whose kernel is $\{1, -1\}$ indicting $SU(2)$ is a double cover of $SO(3)$. In four-dimension, we have similar forms like $r' = q_L r q_R$ where $q_{L,R} \in SU(2)$ and $r$ & $r' \in \mathbb{H}$ are identified with $\vec{r}$ & $\vec{r}' \in \mathbb{R}^4$, respectively. Likewise, a surjective homomorphism $\rho: (q_L, q_R) \in SU(2) \times SU(2) \to R \in SO(4)$ with kernel being $\{(1,1), (-1,-1)\}$ also enables a double-cover. All above properties can be summarized as $Spin(N)$ is a double cover of $SO(N)$, and there are group isomorphisms $Spin(3) \cong SU(2)$ and $Spin(4) \cong SU(2) \times SU(2)$.

Left/right isoclinic rotations are represented by left/right multiplication of unit quaternions. Thus, any rotation in four-dimension can be factorized into the commutative composition of two isoclinic rotations, i.e. $R = R_{q_L} R_{q_R}$. We denote $R_{q_L}(r) = q_L r$ with $q_L = a + bi + cj + dk$ and $r = u + xi + yj + zk$. Then,

$$R_{q_L}(r)=\begin{pmatrix} a & -b & -c & -d \\ b & a & -d & c \\ c & d & a & -b \\ d & -c & b & a \end{pmatrix}\begin{pmatrix} u \\ x \\ y \\ z \end{pmatrix} \tag{S11}$$

This is a left quaternion multiplication of $r$ by $q_L$. For a right quaternion multiplication, i.e. $R_{q_R}(r) = rq_R$. Assuming $q_R = p + qi + rj + sk$, we have,

$$R_{q_R}(r)=\begin{pmatrix} p & -q & -r & -s \\ q & p & s & -r \\ r & -s & p & q \\ s & r & -q & p \end{pmatrix}\begin{pmatrix} u \\ x \\ y \\ z \end{pmatrix} \tag{S12}$$

It is easy to see $R_{q_L}\left(R_{q_R}(r)\right) = R_{q_R}\left(R_{q_L}(r)\right) = q_L r q_R$ where quaternion multiplication is associative. Thus, the two isoclinic rotations are commutative.

# 3. Tight-binding model

The real-space Hamiltonian reads,

$$\mathcal{H} = \sum_n \begin{pmatrix} \sum_{\substack{X=A,B,C,D \\ Y=A,B,C,D}} s_{XY} c_{X,n}^{\dagger} c_{Y,n} + \\ \sum_{\substack{X=A,B,C,D \\ Y=A,B,C,D}} v_{XY} c_{X,n}^{\dagger} c_{Y,n+1} + \sum_{\substack{X=A,B,C,D \\ Y=A,B,C,D}} v_{XYl} c_{X,n}^{\dagger} c_{Y,n+2} + h.c. \end{pmatrix} \tag{S13}$$

where $c_{X,n}^{\dagger}$ and $c_{X,n}$ are creation and annihilation operators on the sub-lattice '$X/Y$' and site '$n$', respectively. Here, we consider a more general case having both the NN (nearest neighbour) and NNN (next-nearest neighbour) hoppings. After Fourier transformation we obtain,

$$H(k) = \begin{bmatrix} s_{AA} + 2v_{AA}\cos k & 2w_{AB}\sin k & 2w_{AC}\sin k & 2w_{AD}\sin k \\ 2w_{AB}\sin k & s_{BB} + 2v_{BB}\cos k & 2w_{BC}\sin k & 2w_{BD}\sin k \\ 2w_{AC}\sin k & 2w_{BC}\sin k & s_{CC} + 2v_{CC}\cos k & 2w_{CD}\sin k \\ 2w_{AD}\sin k & 2w_{BD}\sin k & 2w_{CD}\sin k & s_{DD} + 2v_{DD}\cos k \end{bmatrix}$$
$$+ \begin{bmatrix} 2v_{AAl}\cos 2k & 2w_{ABl}\sin 2k & 2w_{ACl}\sin 2k & 2w_{ADl}\sin 2k \\ 2w_{ABl}\sin 2k & 2v_{BBl}\cos 2k & 2w_{BCl}\sin 2k & 2w_{BDl}\sin 2k \\ 2w_{ACl}\sin 2k & 2w_{BCl}\sin 2k & 2v_{CCl}\cos 2k & 2w_{CDl}\sin 2k \\ 2w_{ADl}\sin 2k & 2w_{BDl}\sin 2k & 2w_{CDl}\sin 2k & 2v_{DDl}\cos 2k \end{bmatrix}$$
$$+ \begin{bmatrix} 0 & s_{AB} + 2r_{AB}\cos k & s_{AC} + 2r_{AC}\cos k & s_{AD} + 2r_{AD}\cos k \\ s_{AB} + 2r_{AB}\cos k & 0 & s_{BC} + 2r_{BC}\cos k & s_{BD} + 2r_{BD}\cos k \\ s_{AC} + 2r_{AC}\cos k & s_{BC} + 2r_{BC}\cos k & 0 & s_{CD} + 2r_{CD}\cos k \\ s_{AD} + 2r_{AD}\cos k & s_{BD} + 2r_{BD}\cos k & s_{CD} + 2r_{CD}\cos k & 0 \end{bmatrix}$$

$$
+\begin{bmatrix} 0 & 2r_{ABl}\cos 2k & 2r_{ACl}\cos 2k & 2r_{ADl}\cos 2k \\ 2r_{ABl}\cos 2k & 0 & 2r_{BCl}\cos 2k & 2r_{BDl}\cos 2k \\ 2r_{ACl}\cos 2k & 2r_{BCl}\cos 2k & 0 & 2r_{CDl}\cos 2k \\ 2r_{ADl}\cos 2k & 2r_{BDl}\cos 2k & 2r_{CDl}\cos 2k & 0 \end{bmatrix} \tag{S14}
$$

where we have set,

$$
\begin{aligned}
v_{AB(l)} &= r_{AB(l)} + i w_{AB(l)} = v_{BA(l)} \\
v_{AC(l)} &= r_{AC(l)} + i w_{AC(l)} = v_{CA(l)} \\
v_{AD(l)} &= r_{AD(l)} + i w_{AD(l)} = v_{DA(l)} \\
v_{BC(l)} &= r_{BC(l)} + i w_{BC(l)} = v_{CB(l)} \\
v_{BD(l)} &= r_{BD(l)} + i w_{BD(l)} = v_{DB(l)} \\
v_{CD(l)} &= r_{CD(l)} + i w_{CD(l)} = v_{DC(l)}
\end{aligned} \tag{S15}
$$

Table S3. Tight-binding coefficients of the ideal flat band models for different non-Abelian topological charges (charge $-1$ needs next nearing neighbour hoppings).

$s_{XX}$ coefficient

| | $q_{12}$ | $-q_{12}$ | $q_{13}$ | $-q_{13}$ | $q_{14}$ | $-q_{14}$ | $q_{23}$ | $-q_{23}$ | $q_{24}$ | $-q_{24}$ | $q_{34}$ | $-q_{34}$ | $q_{1234}$ | $-q_{1234}$ | +1 | -1 |
|---|---|---|---|---|---|---|---|---|---|---|---|---|---|---|---|---|
| $s_{AA}$ | 3/2 | 3/2 | 2 | 2 | 5/2 | 5/2 | 1 | 1 | 1 | 1 | 1 | 1 | 3/2 | 3/2 | 1 | |
| $s_{BB}$ | 3/2 | 3/2 | 2 | 2 | 2 | 2 | 5/2 | 5/2 | 3 | 3 | 2 | 2 | 3/2 | 3/2 | 2 | |
| $s_{CC}$ | 3 | 3 | 2 | 2 | 3 | 3 | 5/2 | 5/2 | 3 | 3 | 7/2 | 7/2 | 7/2 | 7/2 | 3 | |
| $s_{DD}$ | 4 | 4 | 4 | 4 | 5/2 | 5/2 | 4 | 4 | 3 | 3 | 7/2 | 7/2 | 7/2 | 7/2 | 4 | |

$v_{XX}$ coefficient

| | $q_{12}$ | $-q_{12}$ | $q_{13}$ | $-q_{13}$ | $q_{14}$ | $-q_{14}$ | $q_{23}$ | $-q_{23}$ | $q_{24}$ | $-q_{24}$ | $q_{34}$ | $-q_{34}$ | $q_{1234}$ | $-q_{1234}$ | +1 | -1 |
|---|---|---|---|---|---|---|---|---|---|---|---|---|---|---|---|---|
| $v_{AA}$ | 1/4 | 1/4 | 1/2 | 1/2 | 3/4 | 3/4 | 0 | 0 | 0 | 0 | 0 | 0 | 1/4 | 1/4 | 0 | |
| $v_{BB}$ | -1/4 | -1/4 | 0 | 0 | 0 | 0 | 1/4 | 1/4 | 1/2 | 1/2 | 0 | 0 | -1/4 | -1/4 | 0 | |
| $v_{CC}$ | 0 | 0 | -1/2 | -1/2 | 0 | 0 | -1/4 | -1/4 | 0 | 0 | 1/4 | 1/4 | 1/4 | 1/4 | 0 | |
| $v_{DD}$ | 0 | 0 | 0 | 0 | -3/4 | -3/4 | 0 | 0 | -1/2 | -1/2 | -1/4 | -1/4 | -1/4 | -1/4 | 0 | |

$w_{XY}$ coefficient

| | $q_{12}$ | $-q_{12}$ | $q_{13}$ | $-q_{13}$ | $q_{14}$ | $-q_{14}$ | $q_{23}$ | $-q_{23}$ | $q_{24}$ | $-q_{24}$ | $q_{34}$ | $-q_{34}$ | $q_{1234}$ | $-q_{1234}$ | +1 | -1 |
|---|---|---|---|---|---|---|---|---|---|---|---|---|---|---|---|---|
| $w_{AB}$ | 1/4 | -1/4 | 0 | 0 | 0 | 0 | 0 | 0 | 0 | 0 | 0 | 0 | 1/4 | -1/4 | 0 | |
| $w_{AC}$ | 0 | 0 | 1/2 | -1/2 | 0 | 0 | 0 | 0 | 0 | 0 | 0 | 0 | 0 | 0 | 0 | |
| $w_{AD}$ | 0 | 0 | 0 | 0 | 3/4 | -3/4 | 0 | 0 | 0 | 0 | 0 | 0 | 0 | 0 | 0 | |
| $w_{BC}$ | 0 | 0 | 0 | 0 | 0 | 0 | 1/4 | -1/4 | 0 | 0 | 0 | 0 | 0 | 0 | 0 | |
| $w_{BD}$ | 0 | 0 | 0 | 0 | 0 | 0 | 0 | 0 | 1/2 | -1/2 | 0 | 0 | 0 | 0 | 0 | |
| $w_{CD}$ | 0 | 0 | 0 | 0 | 0 | 0 | 0 | 0 | 0 | 0 | 1/4 | -1/4 | 1/4 | 1/4 | 0 | |

Table S4. Tight-binding coefficients of the ideal flat band models for different factorizations of charges $\pm q_{1234}$.

$s_{XX}$ coefficient

| | $q_{1234}=q_{12}q_{34}$ | $-q_{1234}=-q_{12}q_{34}$ | $q_{1234}=-q_{13}q_{24}$ | $-q_{1234}=q_{13}q_{24}$ | $q_{1234}=q_{14}q_{23}$ | $-q_{1234}=-q_{14}q_{23}$ |
|---|---|---|---|---|---|---|
| $s_{AA}$ | 3/2 | 3/2 | 2 | 2 | 5/2 | 5/2 |
| $s_{BB}$ | 3/2 | 3/2 | 3 | 3 | 5/2 | 5/2 |
| $s_{CC}$ | 7/2 | 7/2 | 2 | 2 | 5/2 | 5/2 |
| $s_{DD}$ | 7/2 | 7/2 | 3 | 3 | 5/2 | 5/2 |

$v_{XX}$ coefficient

| | $q_{1234}=q_{12}q_{34}$ | $-q_{1234}=-q_{12}q_{34}$ | $q_{1234}=-q_{13}q_{24}$ | $-q_{1234}=q_{13}q_{24}$ | $q_{1234}=q_{14}q_{23}$ | $-q_{1234}=-q_{14}q_{23}$ |
|---|---|---|---|---|---|---|
| $v_{AA}$ | 1/4 | 1/4 | 1/2 | 1/2 | 3/4 | 3/4 |
| $v_{BB}$ | -1/4 | -1/4 | 1/2 | 1/2 | 1/4 | 1/4 |
| $v_{CC}$ | 1/4 | 1/4 | -1/2 | -1/2 | -1/4 | -1/4 |
| $v_{DD}$ | -1/4 | -1/4 | -1/2 | -1/2 | -3/4 | -3/4 |

$w_{XY}$ coefficient

| | $q_{1234}=q_{12}q_{34}$ | $-q_{1234}=-q_{12}q_{34}$ | $q_{1234}=-q_{13}q_{24}$ | $-q_{1234}=q_{13}q_{24}$ | $q_{1234}=q_{14}q_{23}$ | $-q_{1234}=-q_{14}q_{23}$ |
|---|---|---|---|---|---|---|
| $w_{AB}$ | 1/4 | -1/4 | 0 | 0 | 0 | 0 |
| $w_{AC}$ | 0 | 0 | -1/2 | 1/2 | 0 | 0 |
| $w_{AD}$ | 0 | 0 | 0 | 0 | 3/4 | -3/4 |
| $w_{BC}$ | 0 | 0 | 0 | 0 | 1/4 | 1/4 |
| $w_{BD}$ | 0 | 0 | 1/2 | 1/2 | 0 | 0 |
| $w_{CD}$ | 1/4 | 1/4 | 0 | 0 | 0 | 0 |

Table S5. Integer-valued tight-binding coefficients of the general simulation and experiment models for transmission line networks.

$s_{XX}$ coefficient

| | $-q_{14}$ | -1 | $-q_{1234}=-(q_{12}q_{34})$ | $q_{1234}=(-q_{13}q_{24})$ | $-q_{1234}=-(q_{14}q_{23})$ |
|---|---|---|---|---|---|
| $s_{AA}$ | -4 | -2 | -4 | -2 | 0 |
| $s_{BB}$ | -2 | -2 | -4 | -2 | 0 |
| $s_{CC}$ | 2 | -2 | 4 | 0 | 0 |
| $s_{DD}$ | 4 | 2 | 4 | 0 | 0 |

$v_{XX}$ coefficient

| | $-q_{14}$ | -1 | $-q_{1234}=-(q_{12}q_{34})$ | $q_{1234}=(-q_{13}q_{24})$ | $-q_{1234}=-(q_{14}q_{23})$ |
|---|---|---|---|---|---|
| $v_{AA}$ | 1 | -1 | -1 | 1 | 1 |
| $v_{BB}$ | -2 | 0 | 1 | -1 | -2 |
| $v_{CC}$ | 2 | 1 | 1 | 1 | 2 |
| $v_{DD}$ | -1 | 0 | -1 | -1 | -1 |

$w_{XY}$ coefficient

| | $-q_{14}$ | -1 | $-q_{1234}=-(q_{12}q_{34})$ | $q_{1234}=(-q_{13}q_{24})$ | $-q_{1234}=-(q_{14}q_{23})$ |
|---|---|---|---|---|---|
| $w_{AB}$ | 1 | 1 | 1 | 1 | 1 |
| $w_{AC}$ | 0 | 0 | 0 | 0 | 0 |
| $w_{AD}$ | 0 | 0 | 0 | 0 | 0 |
| $w_{BC}$ | 1 | 1 | 1 | 1 | 1 |
| $w_{BD}$ | 0 | 0 | 0 | 0 | 0 |
| $w_{CD}$ | 1 | 1 | 1 | 1 | -1 |

# 4. Analytical solutions of edge states for the flat-band models

Here we present an analytic method to find the exact solutions of edge states for the flat-band models. We consider the following five types: (i) Edge states of the charges $\pm q_{mn}$; (ii) Edge states of the charges $\pm q_{1234}$; (iii) Evolution of edge states of the charges $\pm q_{1234}$ between different factorizations; (iv) Edge states of the charge $-1$ represented by $q_{mn}^2$; (v) Evolution of edge states of the charge $-1$.

### (i) Edge states of the charges ±qmn

The 1D Hamiltonian corresponding to the above charges can be constructed as $H(k) = R(k)diag(1,2,3,4)R^T(k)$ , where the rotation matrix $R(k) = e^{\frac{k}{2}L_{ij}}$ with $k = 0 \to 2\pi$ and $i,j = 1,2,3,4$. It should be noted that the choice of $e^{\frac{k}{2}L_{ij}}$ is different from the choice of $e^{\frac{k+\pi}{2}L_{ij}}$ with $k = -\pi \to \pi$ (used in the main text). The eigen energy of edge states is independent of the choice. In order to find the edge states of the system, we rewrite $H(k)$ in the form of,

$$H(k) = H_{11} + H_{12}^* e^{-ik} + H_{12} e^{ik} \tag{S16}$$

Take the charge $q_{34}$ as an example. We have,

$$H_{11} = \begin{pmatrix} 1 & 0 & 0 & 0 \\ 0 & 2 & 0 & 0 \\ 0 & 0 & \frac{7}{2} & 0 \\ 0 & 0 & 0 & \frac{7}{2} \end{pmatrix} \tag{S17}$$

and,

$$H_{12}^* = \frac{1}{4}\begin{pmatrix} 0 & 0 & 0 & 0 \\ 0 & 0 & 0 & 0 \\ 0 & 0 & -1 & -i \\ 0 & 0 & -i & 1 \end{pmatrix} \tag{S18}$$

In terms of the tight-binding model, the diagonal elements of $H_{11}$ describe the site energies and $H_{12}^*$ and $H_{12}$ describe the nearest-neighbour hoppings between two sites. Since all bulk modes are strongly localized in real space due to the fact that all bands are flat, it is natural to assume that the edge modes of the system are also strongly localized at the boundaries. For a system of $N$ sites, we use the following ansatz for the edge state wave functions: $\Psi_L = (a,b,c,d)^T \delta_{n,1}$ and $\Psi_R = (a',b',c',d')^T \delta_{n,N}$, for the left and right edge states, respectively. To study the edge state at the left end, we let $N$ goes to infinity. Since the wave function vanishes for all sites except $n = 1$, the only non-trivial equations of motion at sites $n = 1$ and 2 we need to consider are $(H_{11} - E)(a, b, c, d)^T = 0$ and $H_{12}^*(a, b, c, d)^T = 0$, respectively. It is easy to see that the matrix $H_{12}^*$ is defective and there exists only one solution for the edge state: $\Psi_L = (0,0,1, i)^T \delta_{n,1}$ and $E = \frac{7}{2}$. Similarly, for the edge state at the right boundary, we find from the equations of motion at sites $n = N$ and $N - 1$, i.e., $(H_{11} - E)(a', b', c', d')^T = 0$ and $H_{12}(a', b', c', d')^T = 0$, the solution $\Psi_R = (0,0,1, -i)^T \delta_{n,N}$ and $E = \frac{7}{2}$. Thus two edge states are degenerate and have their wave functions in complex conjugate pairs. This is the result of PT symmetry. In the following, we will only consider the left edge state. The method can be equally applied to all other $\pm q_{mn}$ charges in this subsection. For charge $-q_{34}$, we only need to replace $k$ by $-k$ in $H(k)$. Since the solutions found here satisfy the equations of motion for all sites, they are exact solutions. Here we do not concern with the normalization of the wave function.

### (ii) Edge states of the charges ±q1234

This charge $q_{1234}$ can be factorized into three different configurations, i.e., $q_{12}q_{34}$, $-q_{13}q_{24}$ and $q_{14}q_{23}$. The corresponding rotation matrices are $R(k) = e^{kL_{12}/2}e^{kL_{34}/2}$, $e^{-kL_{13}/2}e^{kL_{24}/2}$ and $e^{kL_{14}/2}e^{kL_{23}/2}$, respectively, from which we can obtain $H(k)$ for each configuration. By expressing $H(k)$ in terms of $H_{11}$ and $H_{12}^*$ and solving the equations of motion at the sites $n = 1$ and 2, i.e., $(H_{11} - E)(a, b, c, d)^T = 0$ and $H_{12}^*(a, b, c, d)^T = 0$, we find two edge states for each configuration, i.e., $E = \frac{3}{2}$ with $\Psi_L = (1, i, 0,0)^T \delta_{n,1}$ and $E = \frac{7}{2}$ with $\Psi_L = (0,0, 1, i)^T \delta_{n,1}$ for the case of $q_{12}q_{34}$ ; $E = 2$ with $\Psi_L = (1,0, -i, 0)^T \delta_{n,1}$ and $E = 3$ with $\Psi_L = (0,1, 0, i)^T \delta_{n,1}$ for the case of $-q_{13}q_{24}$; $E = \frac{5}{2}$ with $\Psi_L = (1,0,0, i)^T \delta_{n,1}$ and $E = \frac{5}{2}$ with $\Psi_L = (0,1, i, 0)^T \delta_{n,1}$ for the case of $q_{14}q_{23}$.

### (iii) Evolution of edge states of the charges ±q1234 between different factorizations

Here we first consider the trajectory of edge states when the system is continuously transformed from one configuration to another one. There are three possible cases: (a) From $q_{12}q_{34}$ to $-q_{13}q_{24}$; (b) From $-q_{13}q_{24}$ to $q_{14}q_{23}$; (c) From $q_{14}q_{23}$ to $q_{12}q_{34}$.

For case (c), we consider two commuting hybrid generators $L_a = cos\theta L_{14} + \sin\theta L_{12}$ and $L_b = \cos\theta\, L_{23} + \sin\theta L_{34}$, where the parameter $\theta \in \left[0, \frac{\pi}{2}\right]$ describing the transition from $q_{14}q_{23}$ to $q_{12}q_{34}$. From the corresponding rotation matrix $R(k,\theta) = e^{kL_a/2}e^{kL_b/2}$, we obtain the Hamiltonian $H(k,\theta) = R(k,\theta)diag(1,2,3,4)R^T(k,\theta)$ and the decomposed components:

$$H_{11} = \begin{pmatrix} 2+\frac{1}{2}\cos 2\theta & 0 & \frac{1}{2}\sin 2\theta & 0 \\ 0 & 2+\frac{1}{2}\cos 2\theta & 0 & -\frac{1}{2}\sin 2\theta \\ \frac{1}{2}\sin 2\theta & 0 & 3-\frac{1}{2}\cos 2\theta & 0 \\ 0 & -\frac{1}{2}\sin 2\theta & 0 & 3-\frac{1}{2}\cos 2\theta \end{pmatrix} \tag{S19}$$

$$H_{12}^* = \frac{1}{4}\begin{pmatrix} -2-\cos 2\theta & -i\sin\theta & -\sin 2\theta & -3i\cos\theta \\ -i\sin\theta & -\cos 2\theta & -i\cos\theta & \sin 2\theta \\ -\sin 2\theta & -i\cos\theta & \cos 2\theta & -i\sin\theta \\ -3i\cos\theta & \sin 2\theta & -i\sin\theta & 2+\cos 2\theta \end{pmatrix} \tag{S20}$$

The matrix $H_{12}^*$ is partially defective and $H_{12}^*\varphi = 0$ has two eigenvectors, which can be chosen as $\varphi_1 = (-i, \sin\theta, 0, \cos\theta)^T$ and $\varphi_2 = (-\sin\theta, -i, \cos\theta, 0)^T$. We write the edge state wave functions as $c_1\varphi_1 + c_2\varphi_2$ and solve the equation $(H_{11} - E)(c_1\varphi_1 + c_2\varphi_2) = 0$ for $E$ and $c_{1,2}$. We find the two edge states at energies $E^\pm = \frac{5}{2} \pm \sin\theta$. This result is shown in Fig. 3f. The corresponding wave functions are $\Psi_L^\pm = [\mp i\sec\theta\,\varphi_1 - \sec\theta\,\varphi_2]\delta_{n,1}$. The edge state energies can also be obtained directly from the eigenvalues of $H_{11}$ without knowing the edge state wave functions. Similarly, we apply the same procedure to cases (a) and (b) and find the following trajectories of the edge state energies: $E^\pm = \frac{5}{2} \pm \frac{\sqrt{2}}{4}\sqrt{5+3\cos 2\theta}$ for case (a) and $E^\pm = \frac{5}{2} \pm \frac{1}{2}\cos\theta$ for case (b). These results are plotted in Figs. 3d and 3e, respectively.

## (iv) Edge states of the charge -1 represented by $q_{mn}^2$

For the above cases, the rotation matrix becomes $R(k) = e^{kL_{ij}}$. We can simply replace $k$ in $H(k)$ by $2k$ and rewrite it in the form of $H(k) = H_{11} + H_{13}^*e^{-2ik} + H_{13}e^{2ik}$. In the real space, such replacement represents a next-nearest neighbor hopping. For a finite chain with even number of sites, the tight-binding Hamiltonian becomes two disconnected sublattices, one with indices $n = 1,3,5 \ldots N-1$ and the other with $n = 2,4,6 \ldots N$. For the case of $q_{34}^2$, the matrices $H_{11}$ and $H_{13}^*$ are identical to those shown in Eqs. (S17) and (S18), respectively. From which we find one edge state at the left boundary per sublattice, i.e., $\Psi_{L1} = (0,0,1,i)^T\delta_{n,1}$ and $\Psi_{L2} = (0,0,1,i)^T\delta_{n,2}$ with the same eigen energy at $E = \frac{7}{2}$.

### (v) Evolution of edge states of the charge -1

Here we consider the following three cases: (a) Transition from $q_{ij}^2$ to $q_{jk}^2$; (b) Edge state evolution involving the mixing of $q_{ij}^2$, $q_{jk}^2$ and $q_{ik}^2$ with rotation of eigenvectors in three bands; (c) Edge state evolution involving the mixing of $q_{ij}^2$, $q_{jk}^2$ and $q_{jl}^2$ with rotation of eigenvectors in four bands.

(a) Transition from $q_{ij}^2$ to $q_{jk}^2$

As an example, we consider a continuous transition from $q_{23}^2$ to $q_{12}^2$. We choose a hybrid generator $L = \cos t\, L_{23} + \sin t\, L_{12}$ with $t \in \left[0, \frac{\pi}{2}\right]$. We obtain the Hamiltonian through $H(k,t) = R(k,t) diag(1,2,3,4) R^T(k,t)$ and rewrite it in the form of,

$$H(k,t) = H_{11}(t) + H_{12}^*(t)e^{-ik} + H_{12}(t)e^{ik} + H_{13}^*(t)e^{-2ik} + H_{13}(t)e^{2ik} \quad \text{(S21)}$$

where,

$$H_{11}(t) = \frac{1}{8}\begin{pmatrix} -2\,cos(2t) - 3\,cos(4t) + 13 & 0 & -3\,sin(4t) & 0 \\ 0 & 4(cos(2t)+4) & 0 & 0 \\ -3\,sin(4t) & 0 & -2\,cos(2t) + 3\,cos(4t) + 19 & 0 \\ 0 & 0 & 0 & 32 \end{pmatrix} \quad \text{(S22)}$$

$$H_{12}^* = -\frac{\sin(2t)}{2}\begin{pmatrix} \sin(2t) & i\cos(t) & -\cos(2t) & 0 \\ i\cos(t) & 0 & i\sin(t) & 0 \\ -\cos(2t) & i\sin(t) & -\sin(2t) & 0 \\ 0 & 0 & 0 & 0 \end{pmatrix} \quad \text{(S23)}$$

$$H_{13}^* = \frac{\cos(2t)}{4}\begin{pmatrix} \sin^2(t) & i\sin(t) & -\cos(t)\sin(t) & 0 \\ i\sin(t) & -1 & -i\cos(t) & 0 \\ -\cos(t)\sin(t) & -i\cos(t) & \cos^2(t) & 0 \\ 0 & 0 & 0 & 0 \end{pmatrix} \quad \text{(S24)}$$

At $t = 0$ or $\frac{\pi}{2}$, $H_{12} = 0$ and the system reduces to two disconnected sublattices and $H_{13}$ describes the hopping elements within each sublattice. For a general value of $t$, $H_{12}$ couples two sublattices and it is natural to assume that an edge state occupies two sites at boundary, one from each sublattice. Thus, we use the edge state ansatz with the form of $\Psi_L = \Phi_1\delta_{n,1} + \Phi_2\delta_{n,2}$, where $\Phi_1 = (a,b,c,d)^T$ and $\Phi_2 = (e,f,g,h)^T$. Both the wave function and the edge state energy $E$ are to be determined through solving the equations of motion. Since the wave function vanishes for sites with $n > 2$, we only need to consider the equations of motion of the four boundary sites, i.e.,

$$(H_{11} - E)\Phi_1 + H_{12}\Phi_2 = 0 \quad (n = 1) \quad \text{(S25)}$$

$$(H_{11} - E)\Phi_2 + H_{12}^*\Phi_1 = 0 \quad (n = 2) \quad \text{(S26)}$$

$$H_{13}^*\Phi_1 + H_{12}^*\Phi_2 = 0 \quad (n = 3) \quad \text{(S27)}$$

$$H_{13}^*\Phi_2 = 0 \quad (n = 4) \quad \text{(S28)}$$

Both the matrices $H_{13}^*$ and $H_{12}^*$ are defective. $H_{13}^*$ has two independent eigen vectors, which can be chosen as $(\sin t, i, -\cos t, 0)^T$ and $(\cos t, 0, \sin t, 0)^T$. However, the matrix $H_{12}^*$ has only one coalesced eigenvector in the form of $(\sin t, i, -\cos t\,, 0)^T$. With these eigen vectors of $H_{13}^*$ and $H_{12}^*$, a simple edge state can be obtained immediately by choosing $\Phi_2 = 0$ and $\Phi_1 = (\sin t, i, -\cos t\,, 0)^T$, which satisfies Eqs. S25-28. Eigenvalue of such an edge state is determined from Eq. (S25). From the equation $(H_{11} - E)\Phi_1 = 0$, we obtain the first solution of the edge state,

$$E^0 = 2 + \frac{1}{2}\cos 2t \tag{S29}$$

with,

$$\Psi_L^0 = (\sin t, i, -\cos t\,, 0)^T \delta_{n,1} \tag{S30}$$

The other edge state solutions can be obtained by choosing $\Phi_2 = (\sin t, i, -\cos t, 0)^T$ and $\Phi_1 = c(\cos t, 0, \sin t\,, 0)^T$ so that Eqs. (S27) and (S28) are satisfied automatically. The unknown function $c$ and the eigenvalue $E$ are to be determined by solving Eqs. (S25) and (S26), from which we find two more edge states:

$$E^{\pm} = \frac{1}{8}(16 - 2\cos 2t \pm \sqrt{2}\sqrt{17 + \cos 4t}) \tag{S31}$$

with,

$$\Psi_L^{\pm} = c_{\pm}(\cos t, 0, \sin t\,, 0)^T \delta_{n,1} + (\sin t, i, -\cos t\,, 0)^T \delta_{n,2} \tag{S32}$$

where $c_{\pm} = \frac{-4\sin 2t}{3\cos 2t \pm \sqrt{9 - \sin^2 2t}}$. Eqs. (S29) and (S31) are plotted in the Panel "From $q_{23}^2$ to $q_{12}^2$" of Fig. S10c with $\theta_{23\to 12} = t$.

(b) Edge state evolution involving the mixing of $q_{ij}^2$, $q_{jk}^2$ and $q_{ik}^2$ with rotation of eigenvectors in three bands

Now we consider a more general case where the edge states form surfaces in a 2D parameter space. As an example, we consider a hybrid generator of the form:

$$L = \cos t\, L_{23} + \sin t \cos f\, L_{12} + \sin t \sin f\, L_{13} \tag{S33}$$

where the parameters $t, f \in \left[0, \frac{\pi}{2}\right]$. From the rotation matrix $R(k,t) = e^{kL}$, we obtain the Hamiltonian $H(k,t,f)$ and the decomposed components of $H_{11}$, $H_{12}^*$ and $H_{13}^*$. By solving Eqs. (S25-S28) we can obtain both the edge state energies and wave functions. However, the edge state energies can also be obtained from the roots of the determinant in Eqs. (S25) and (S26), i.e., $\det\begin{pmatrix} H_{11} - EI_4 & H_{12} \\ H_{12}^* & H_{11} - EI_4 \end{pmatrix} = 0$. From which, we find four pairs of degenerate eigen energies. Apart from the trivial one at $E = 4$, the other three are edge state energies of the left and right edge states with the forms of,

$$E^0 = \frac{1}{16}(\cos(2f - 2t) + \cos(2f + 2t) - 2\cos(2f) + 6\cos(2t) + 34) \tag{S34}$$

$$E^{\pm} = \frac{1}{32}\left(\frac{-\cos(2f - 2t) - \cos\big(2(f + t)\big) + 2\cos(2f) - 6\cos(2t) + 62 \pm}{\sqrt{8\cos(4f)\sin^4(t) - 48\cos(2f)\sin^2(t)(\cos(2t) - 5) + 148\cos(2t) + 19\cos(4t) + 409}}\right) \tag{S35}$$

Eqs. (S34) and (S35) are plotted in the Fig. S11a, which shows three surfaces of edge states involving three bands only. Since these edge states do not involve the fourth band, Fig. S11a also describes the edge state evolution of the charge $-1$ in the 3-band model[22] involving the mixtures of factorizations $i^2$, $j^2$and $k^2$.

It should be noted that Fig. S11a shows the edge state surfaces inside the triangle formed by $q_{12}^2$, $q_{23}^2$ and $q_{13}^2$ shown in Figs. S10a and b. At $f = 0$, Eqs. (S34) and (S35) reduce to Eqs. (S29) and (S31), describing the edge state evolution from $q_{23}^2$ to $q_{12}^2$. At $t = \frac{\pi}{2}$, the parameter $f$ describes the evolution from $q_{12}^2$ to $q_{13}^2$ as can be seen from Eq. (S33). The above results reduce to,

$$E^0 = \frac{1}{4}(7 - \cos(2f)) \tag{S36}$$

$$E^{\pm} = \frac{1}{8}\left(17 + \cos(2f) \pm \sqrt{2}\cos f\sqrt{\cos(2f) + 17}\right) \tag{S37}$$

Eqs. (S36) and (S37) are plotted in the Panel "From $q_{12}^2$ to $q_{13}^2$" of Fig. S10c with $\theta_{12\to 13} = f$.

At $f = \frac{\pi}{2}$, the parameter $t$ describes the evolution from $q_{23}^2$ to $q_{13}^2$ as can be seen from Eq. (S33). Eqs. (S34) and (S35) reduce to,

$$E^0 = \frac{1}{4}(9 + \cos(2t)) \tag{S38}$$

$$E^{\pm} = \frac{1}{8}\left(15 - \cos(2t) \pm \sqrt{2}\cos t\sqrt{\cos(2t) + 17}\right) \tag{S39}$$

Eqs. (S38) and (S39) are plotted in the Panel "From $q_{23}^2$ to $q_{13}^2$" of Fig. S10c with $\theta_{23\to 13} = t$. It should be noted the same analytic method can be applied to all edges of the regular octahedron as shown in Figs. S10a and b. It can also be used to obtain the 3D maps of edge state surfaces resulting from the mixing of $q_{ij}^2$, $q_{jk}^2$ and $q_{ik}^2$ for any three bands $i, j$ and $k$.

(c) Edge state evolution involving the mixing of $q_{ij}^2$, $q_{jk}^2$ and $q_{jl}^2$ with rotation of eigenvectors in four bands

Finally, we consider a hybrid generator of the form,

$$L = \cos t\, L_{12} + \sin t\cos f\, L_{13} + \sin t\sin f\, L_{14} \tag{S40}$$

where $t, f \in \left[0, \frac{\pi}{2}\right]$. Different from case (b), Eq. (S40) now produces the edge state surfaces inside the triangle formed by $q_{12}^2$, $q_{13}^2$ and $q_{14}^2$ highlighted in the Figs. S10a and b, in which all four bands are involved. By using the same procedure of case (b), we obtain the four edge state surfaces as shown in Fig. S11b, which fits well with the numerical results shown in Fig. S10d. One of the four surfaces has the form of,

$$E^0 = \frac{1}{16}\left(\cos\left(2(f - t)\right) + \cos\left(2(f + t)\right) - 2\cos(2f) - 6\cos(2t) + 30\right) \tag{S41}$$

The other three ones are complex as the roots of a cubic equation.

Thus, the method can be used to obtain the edge states inside every face triangle of the regular octahedron as shown in the Figs. S10a and b. The number of edge state surfaces depends on the number of bands involved.

# 5. Experimental methods

There are four meta-atoms A, B, C and D in one unit-cell. The hoppings between two meta-atoms are realized by connecting 2-m-long coaxial cables (model: RG58C/U). To achieve the complex hoppings, we create a hidden dimension by placing four nodes in each meta-atom so that four subspaces are allowed. Due to periodic connections in this hidden dimension, the four subspaces correspond to four pseudo angular momenta that are $\exp(i4\varphi_n) = 1$, with $\varphi_1 = 0, \varphi_2 = \frac{\pi}{2}, \varphi_3 = \pi \; and \; \varphi_4 = -\pi/2$. Through the specific excitation from a 4-channel signal generator (Keysight M3201A), we carried out our experiments in the $\varphi_2 = \pi/2$ subspace. The amplitude and phase of voltage of each meta-atom are probed by an oscilloscope (Keysight DSOX2002A). After subsequent Fourier transformation, we obtain the energy bands and eigenstates in the momentum space. Figure S13 shows the specific transmission line network corresponding to charges $\pm q_{14}$.

# 6. Supplementary Figures

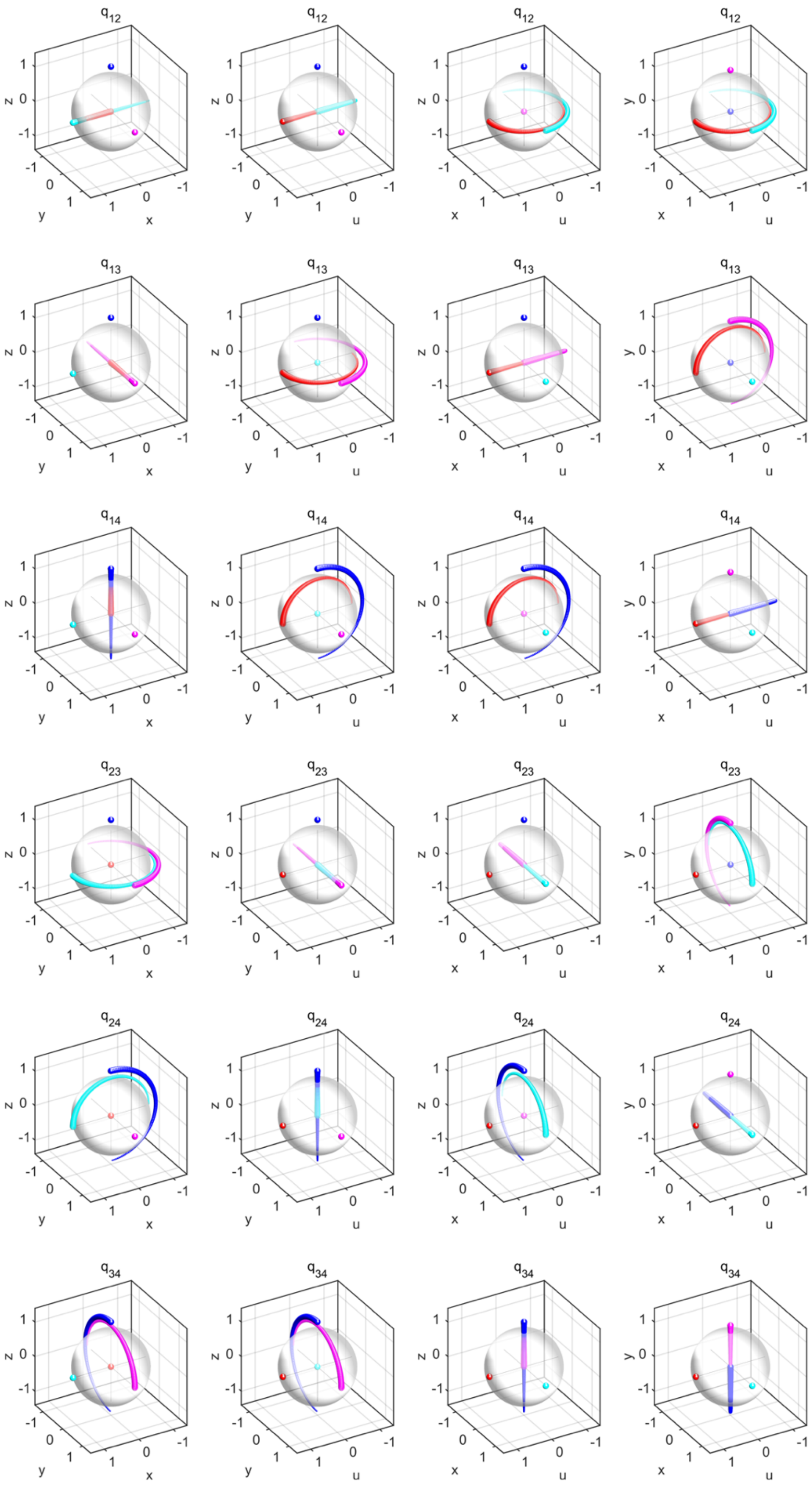
q12 q12 q12 q12
q13 q13 q13 q13
q14 q14 q14 q14
q23 q23 q23 q23
q24 q24 q24 q24
q34 q34 q34 q34
z
y
x
u
1
0
-1

Figure S1. Trajectories of eigenstates of charges $q_{mn}$ orthographically projected onto four solid spheres in $\mathbb{R}^3$. The colours (red, cyan, magenta, blue) correspond to the (first, second, third, fourth) bands. The direction of line-width decreasing indicates $k = -\pi \to \pi$.

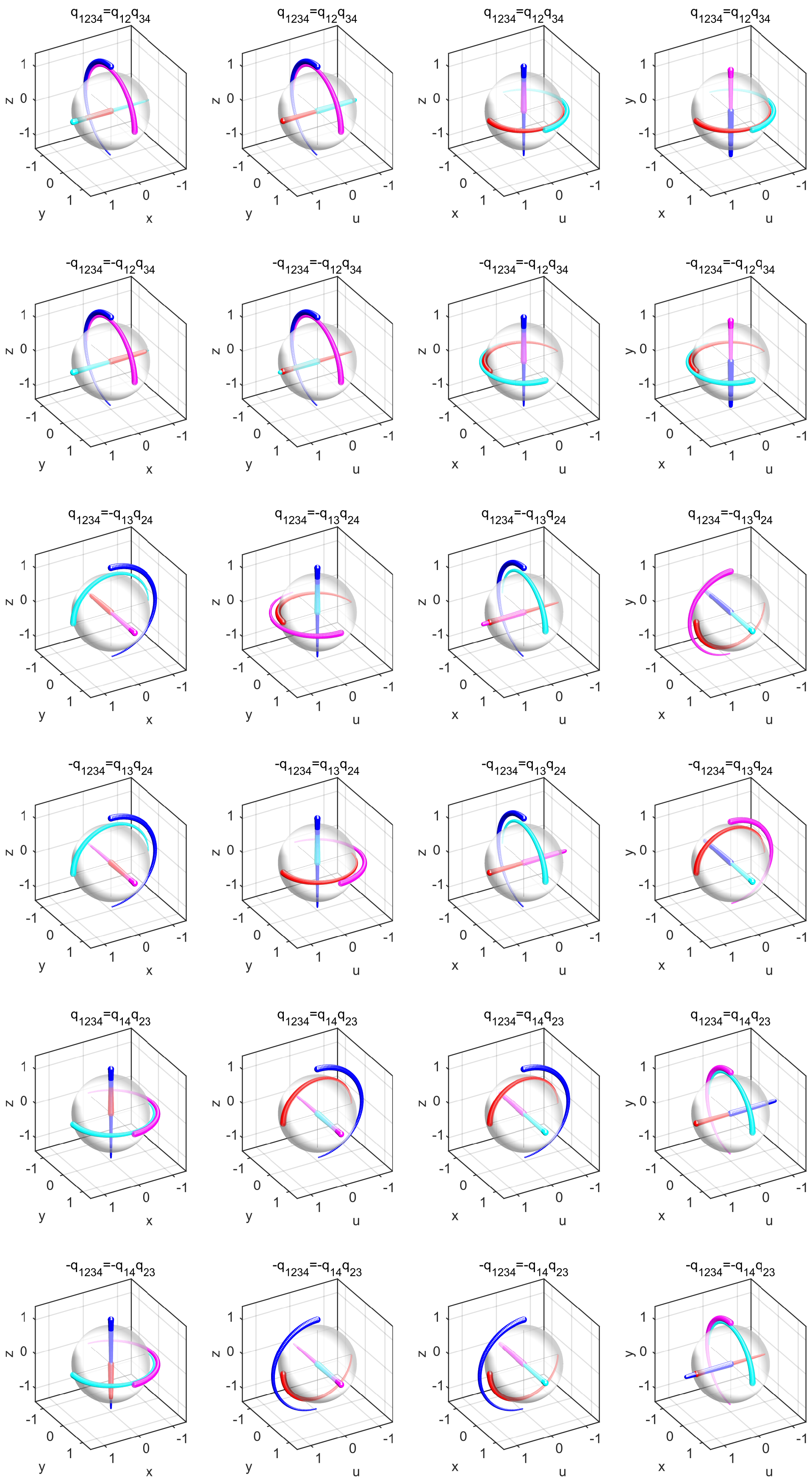

Figure S2. Trajectories of eigenstates of charges $\pm q_{1234}$ orthographically projected onto four solid spheres in $\mathbb{R}^3$. We factorize the charge $q_{1234}$ into three configurations: $q_{1234} = q_{12}q_{34}$, $q_{1234} = -q_{13}q_{24}$ and $q_{1234} = q_{14}q_{23}$. The colours (red, cyan, magenta, blue) correspond to the (first, second, third, fourth) bands. The direction of line-width decreasing indicates $k = -\pi \rightarrow \pi$.

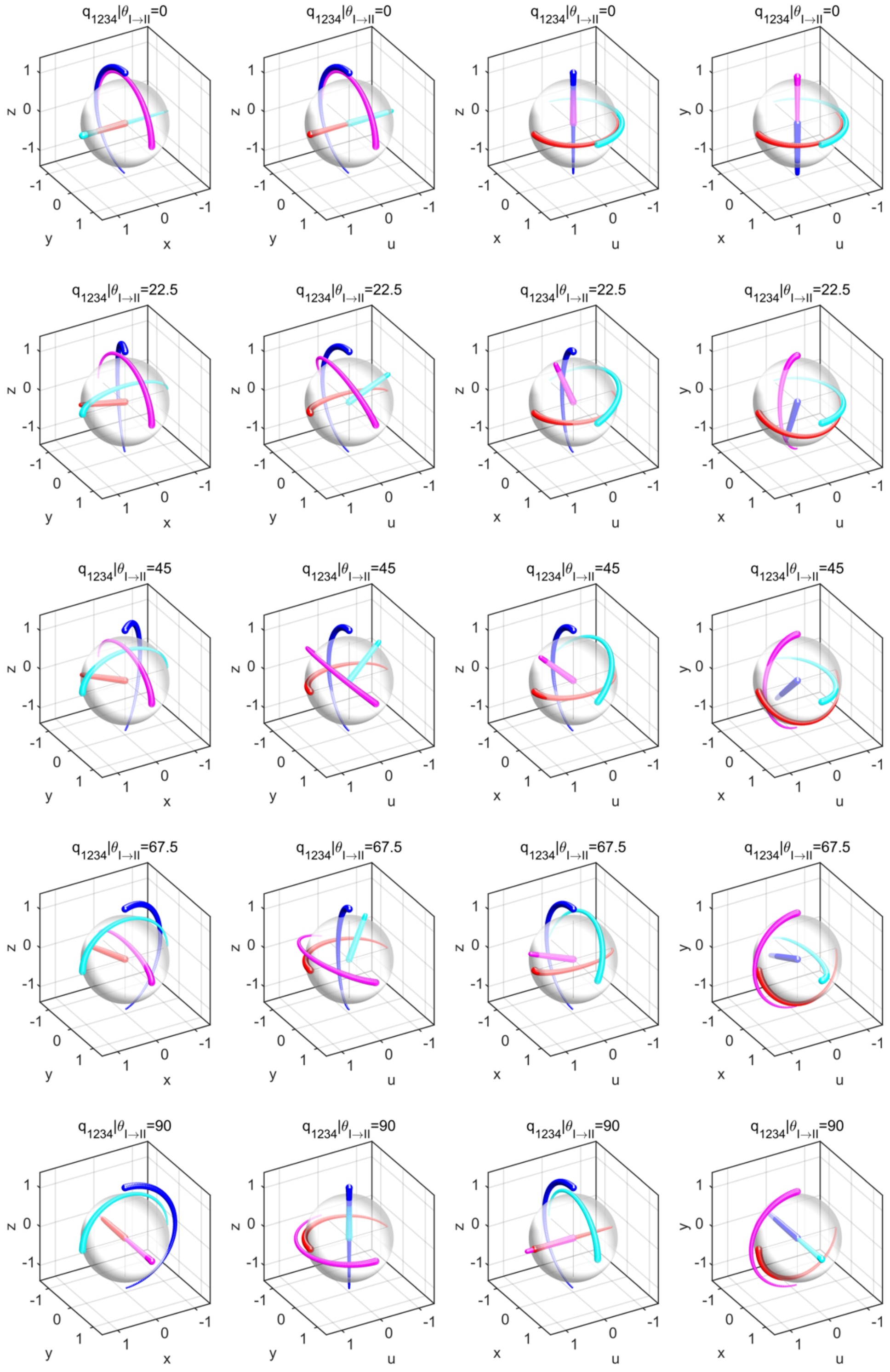

Figure S3. Trajectories of eigenstates of charge $q_{1234}$ orthographically projected onto four solid spheres in $\mathbb{R}^3$. We continuously rotate the charge $q_{1234}$ from the configuration $q_{1234} = q_{12}q_{34}$ to $q_{1234} = -q_{13}q_{24}$, where the parameter $\theta_{I\to II}$ is defined as, $L_a = \cos\theta_{I\to II}\, L_{12} - \sin\theta_{I\to II}\, L_{13}$ *and* $L_b = \cos\theta_{I\to II}\, L_{34} + \sin\theta_{I\to II}\, L_{24}$. The Hamiltonian can be written as $H = \exp\left(\frac{k_\pi}{2} L_b\right)\exp\left(\frac{k_\pi}{2} L_a\right) I_{1234} \exp\left(-\frac{k_\pi}{2} L_a\right)\exp\left(-\frac{k_\pi}{2} L_b\right)$ with $k_\pi = k + \pi$. The colours (red,

cyan, magenta, blue) correspond to the (first, second, third, fourth) bands. The direction of line-width decreasing indicates $k = -\pi \to \pi$.

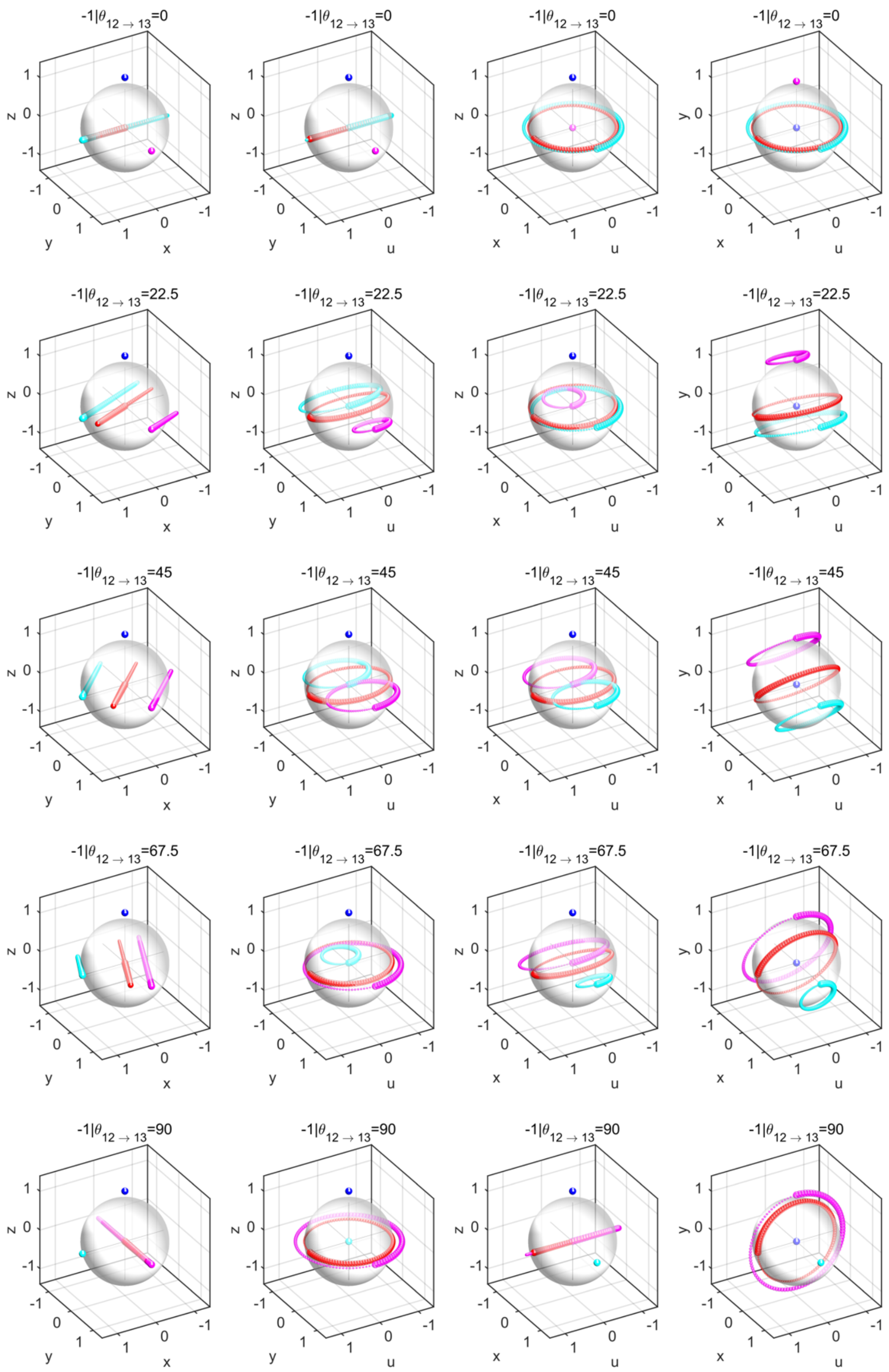

Figure S4. Trajectories of eigenstates of charge $-1$ orthographically projected onto four solid spheres in $\mathbb{R}^3$. We continuously rotate the charge $-1$ from the configuration $-1 = q_{12}^2$ to $-1 = -q_{13}^2$, where the parameter $\theta_{12\to13}$ is defined as, $L = \cos\theta_{12\to13} L_{12} + \sin\theta_{12\to13} L_{13}$. The Hamiltonian can be written as $H = \exp(k_\pi L)\, I_{1234} \exp(-k_\pi L)$ with $k_\pi = k + \pi$. The colours (red, cyan, magenta, blue) correspond to the (first, second, third, fourth) bands. The direction of line-width decreasing indicates $k = -\pi \to \pi$.

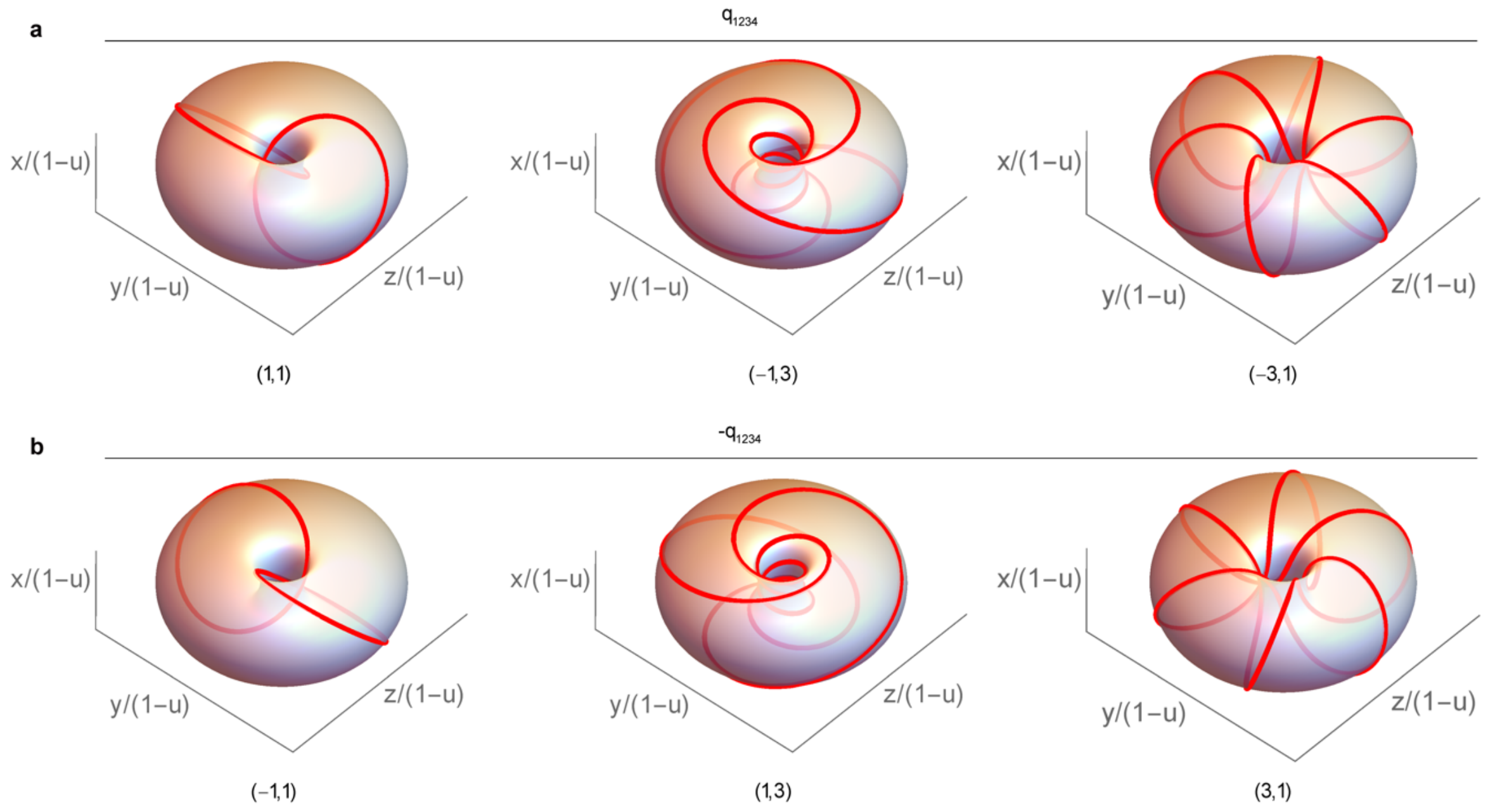

Figure S5. Stereographically projected Clifford tori in $\mathbb{R}^3$, i.e. $(u,x,y,z) \to \left(\frac{x}{1-u}, \frac{y}{1-u}, \frac{z}{1-u}\right)$. The index $(m,n)$ indicates the eigenstates rotate $m\pi$ and $n\pi$ on the $oux$ and $oyz$ planes, respectively. a, The three cases corresponding to charge $+q_{1234}$, they can be continuously transformed into each other. b, The opposite rotation senses corresponding to charge $-q_{1234}$.

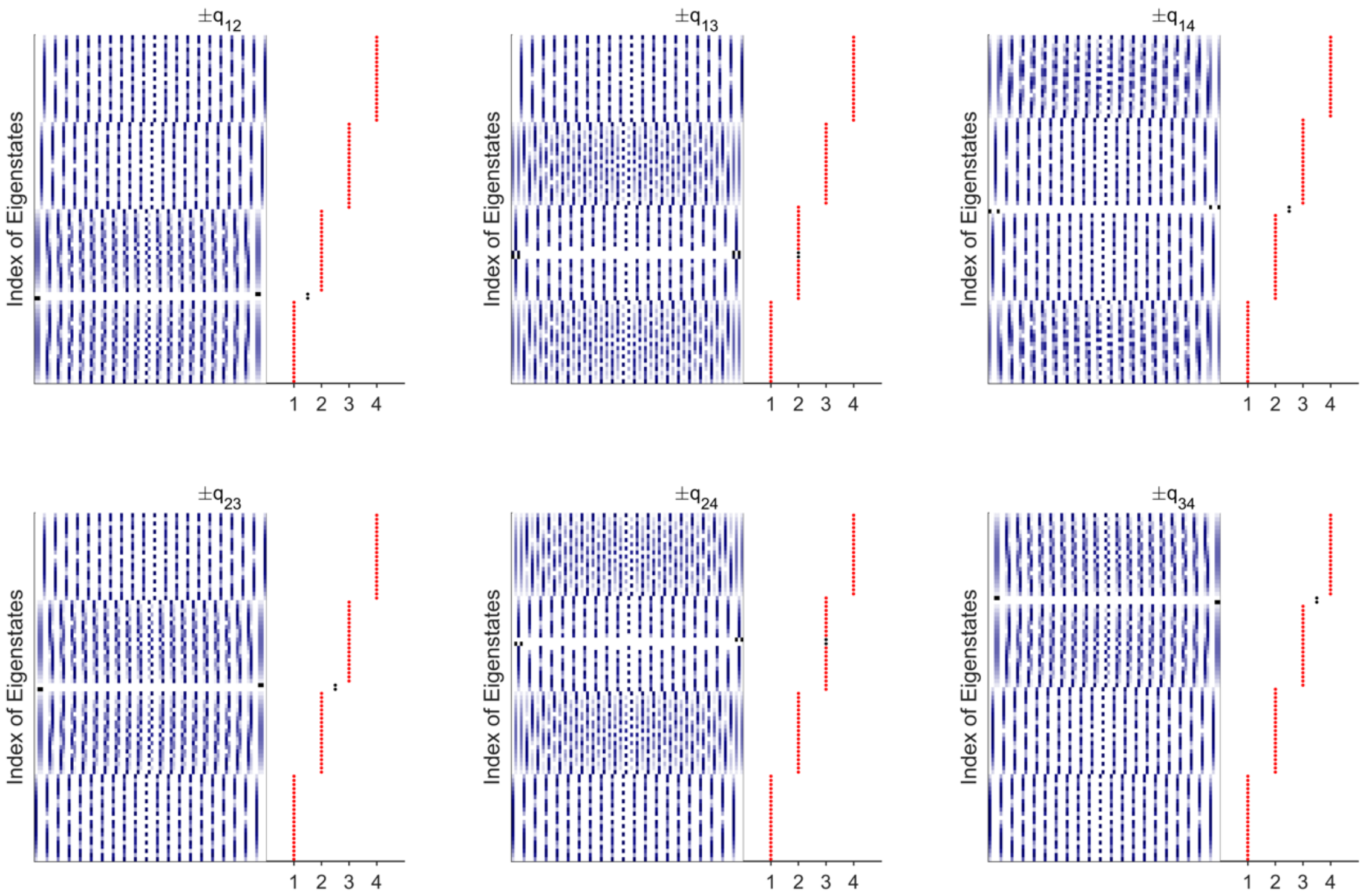

Figure S6. Edge state distributions at the hard boundaries of a finite lattice for charges $\pm q_{mn}$ of flat band models.

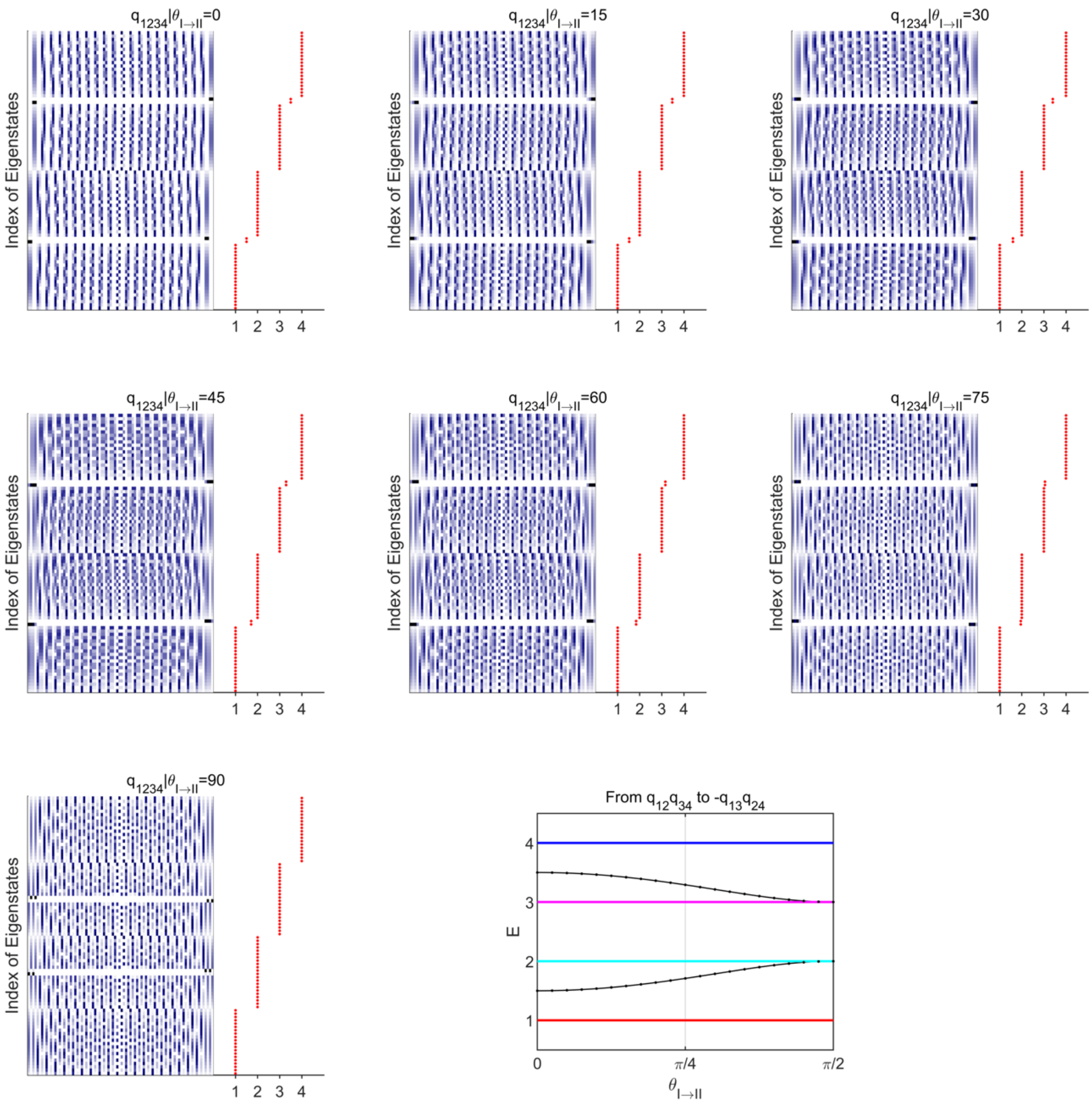

Figure S7. Evolution of edge state distributions for charge $q_{1234}$ from the factorization of $q_{1234} = q_{12}q_{34}$ to $q_{1234} = -q_{13}q_{24}$, parametrized by $\theta_{I\rightarrow II}$ with unit of degrees. Lines/dots indicate numerical/analytical results.

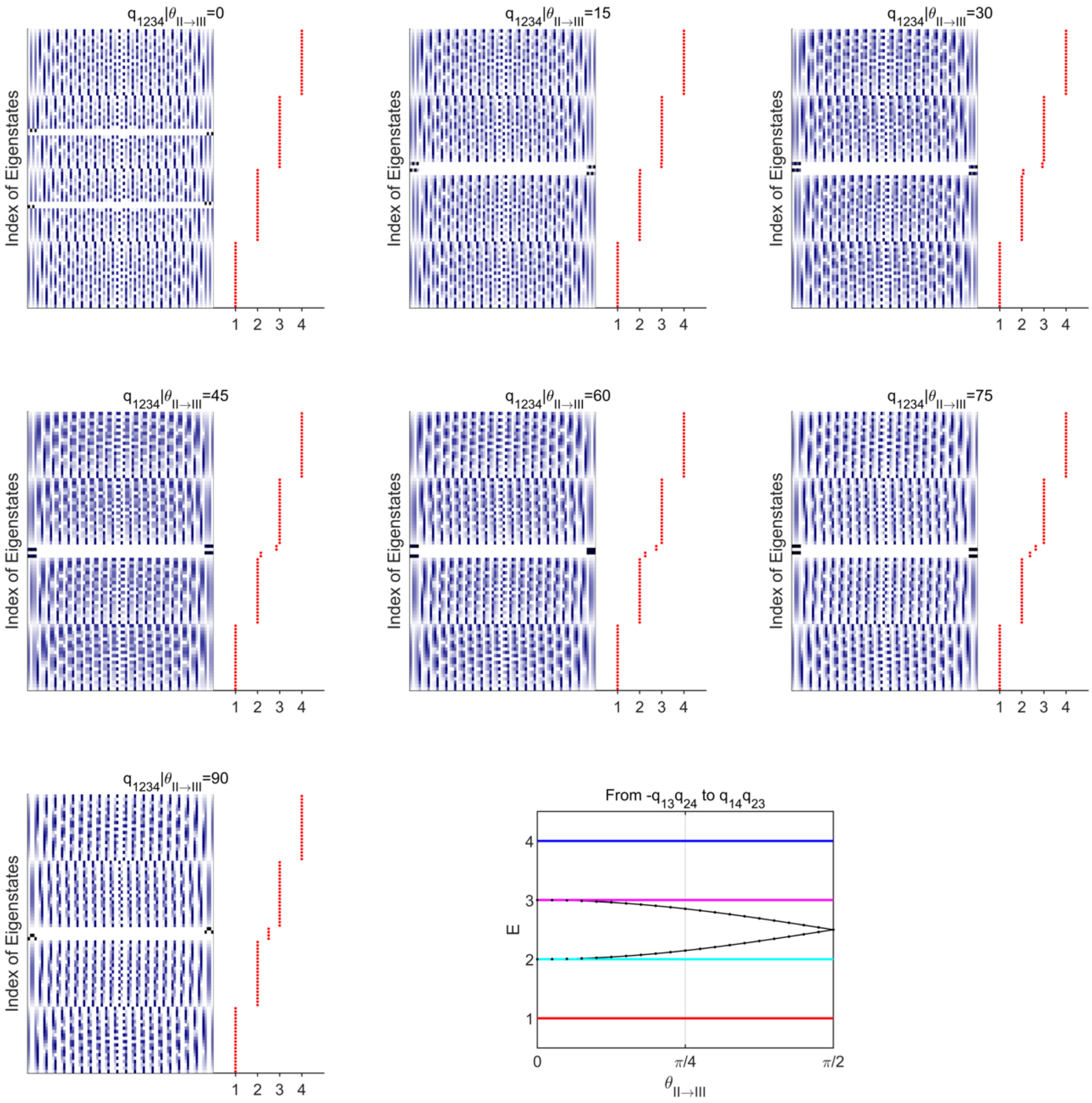

Figure S8. Evolution of edge state distributions for charge $q_{1234}$ from the factorization of $q_{1234} = -q_{13}q_{24}$ to $q_{1234} = q_{14}q_{23}$, parametrized by $\theta_{II\to III}$ with unit of degrees. Lines/dots indicate numerical/analytical results.

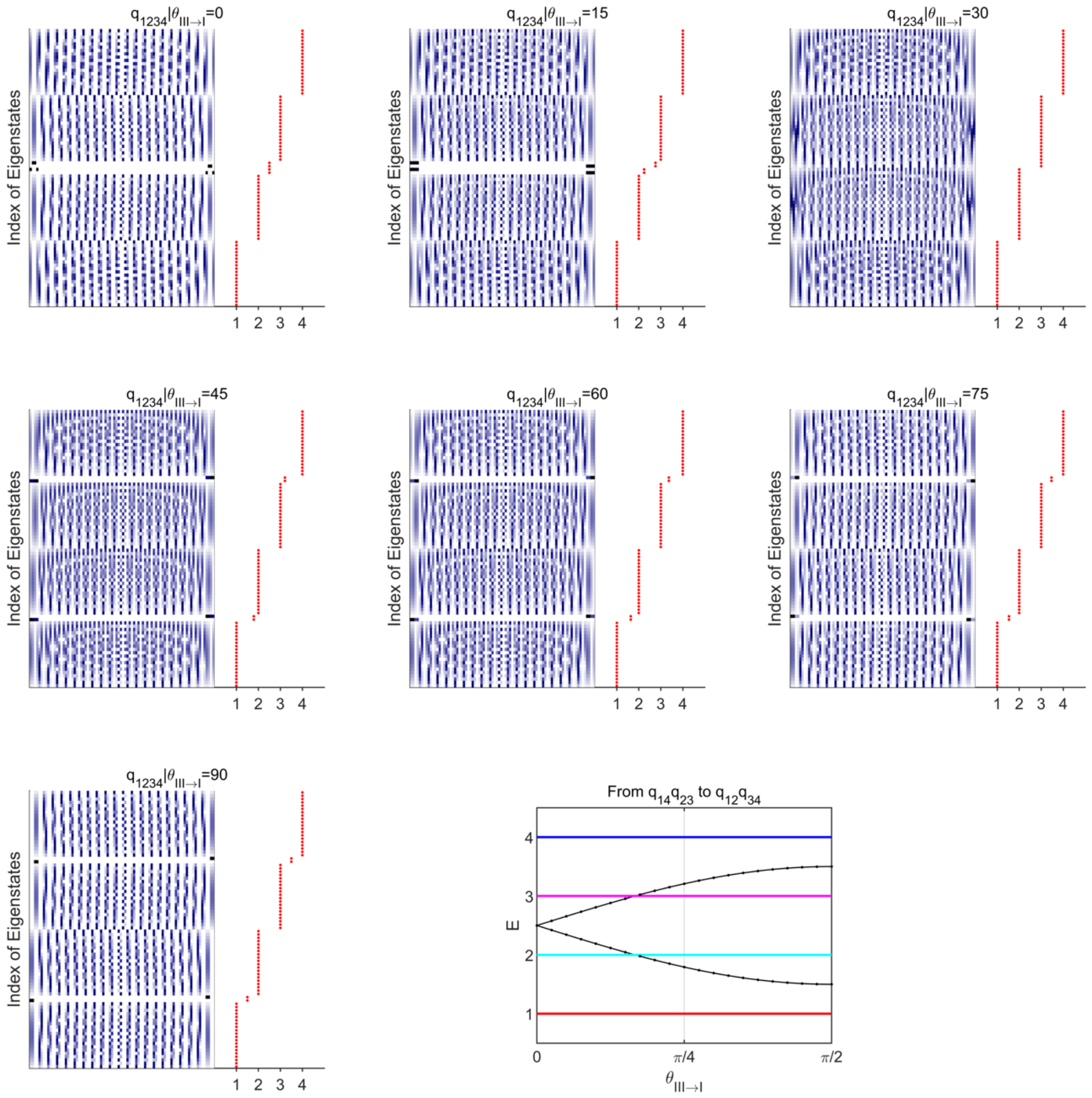

Figure S9. Evolution of edge state distributions for charge $q_{1234}$ from the factorization of $q_{1234} = q_{14}q_{23}$ to $q_{1234} = q_{12}q_{34}$, parametrized by $\theta_{III\to I}$ with unit of degrees. Lines/dots indicate numerical/analytical results.

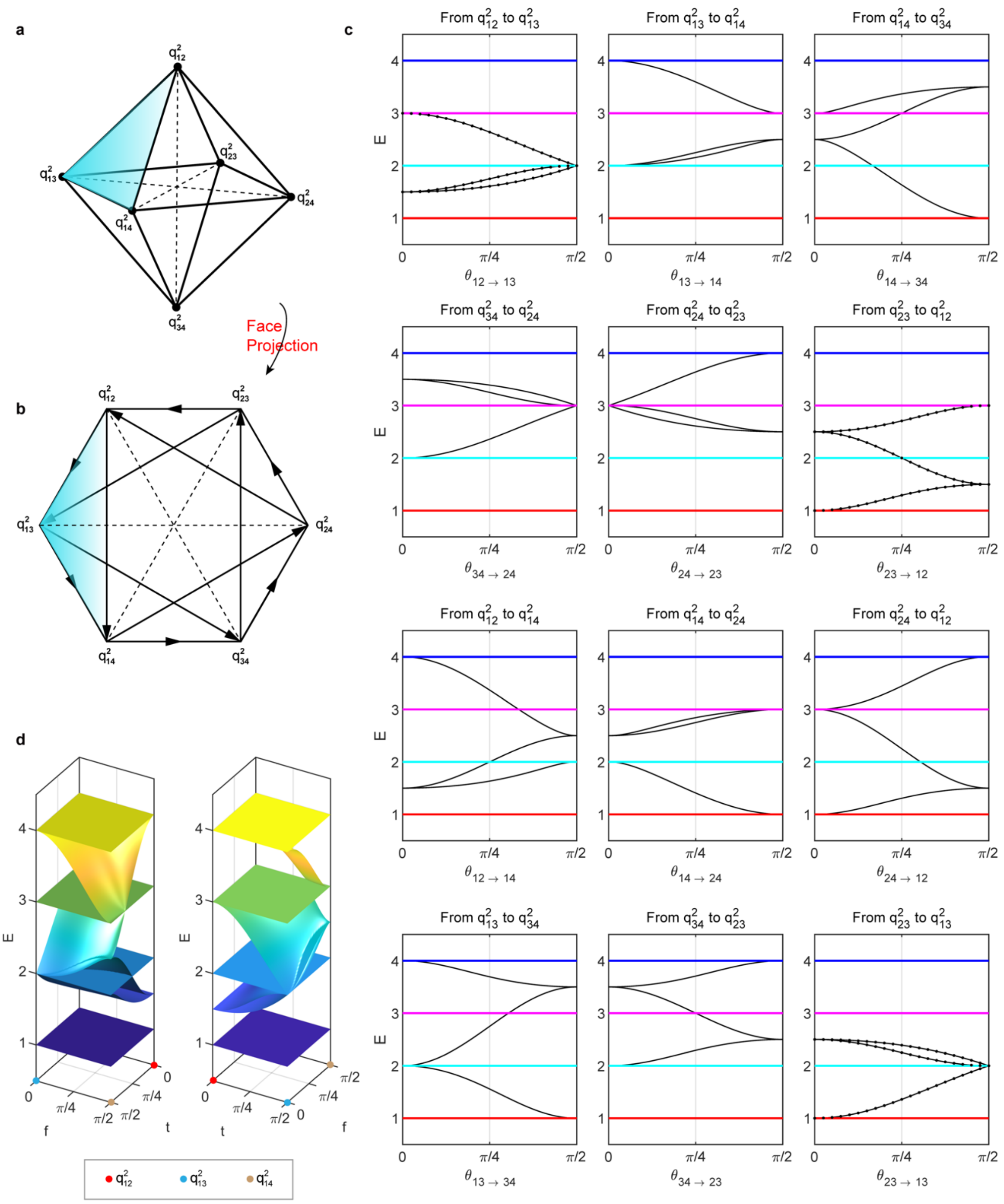

Figure S10. Evolution of edge state distributions for charge $-1$. **a,** All possible factorizations of charge $-1$ illustrated on a regular octahedron. **b,** Orthogonal projection centred by face. There are 12 possible transitions, the direct transition (dashed lines) between one pair of diagonal points is not allowed as they are located on two orthogonal planes, i.e. $q_{12}^2 \nLeftrightarrow q_{34}^2$. **c,** Evolution of edge state distributions along 12 edges of the regular octahedron. **d,** Evolution of edge state distributions on one face (with vertices $q_{12}^2, q_{13}^2$ *and* $q_{14}^2$) of the regular octahedron. Lines/dots indicate numerical/analytical results.

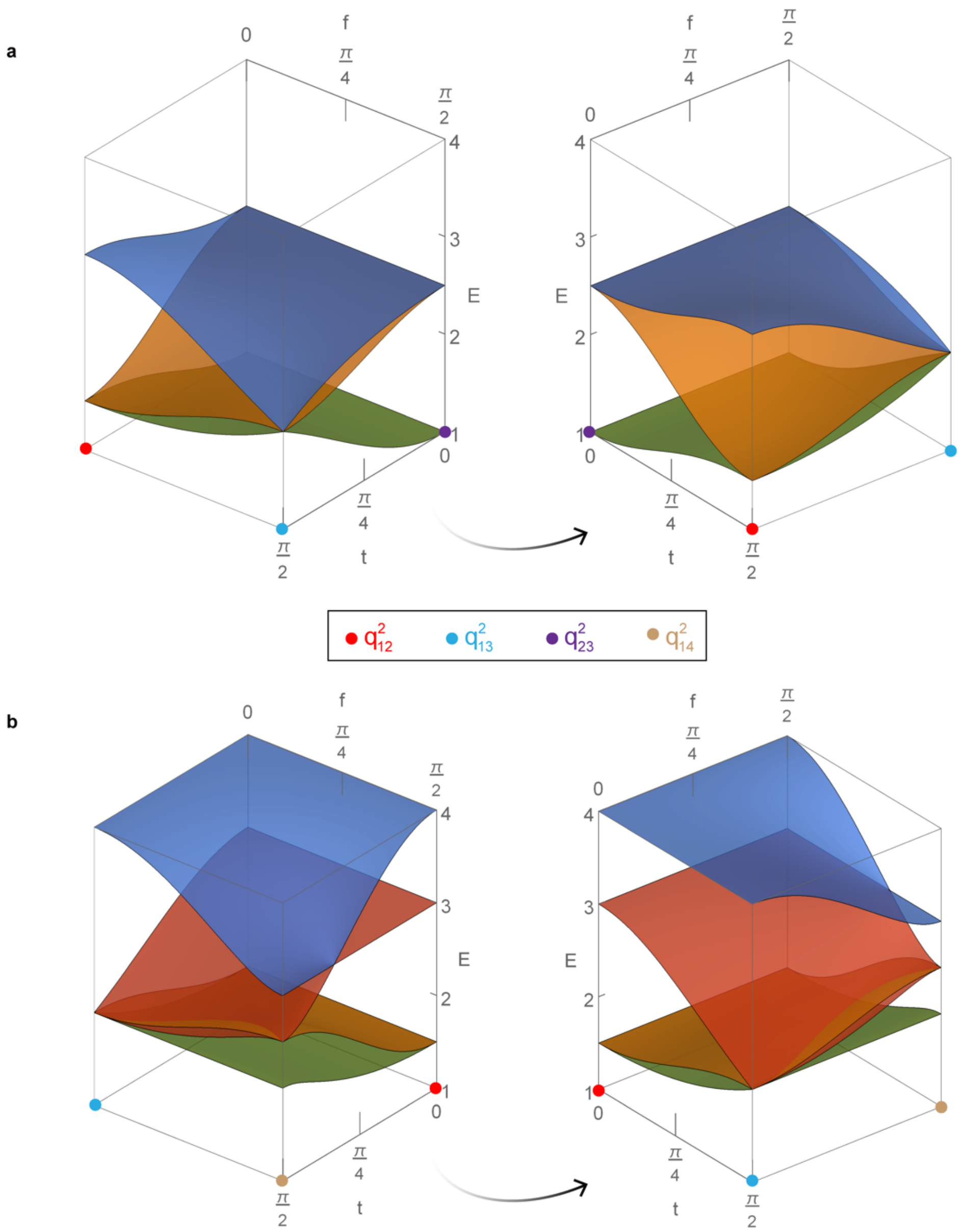

Figure S11. Analytical edge state surfaces. a, Edge state surfaces on the triangle face of $(q_{12}^2, q_{13}^2, q_{23}^2)$ involving three bands. b, Edge state surfaces on the triangle face of $(q_{12}^2, q_{13}^2, q_{14}^2)$ involving four bands.

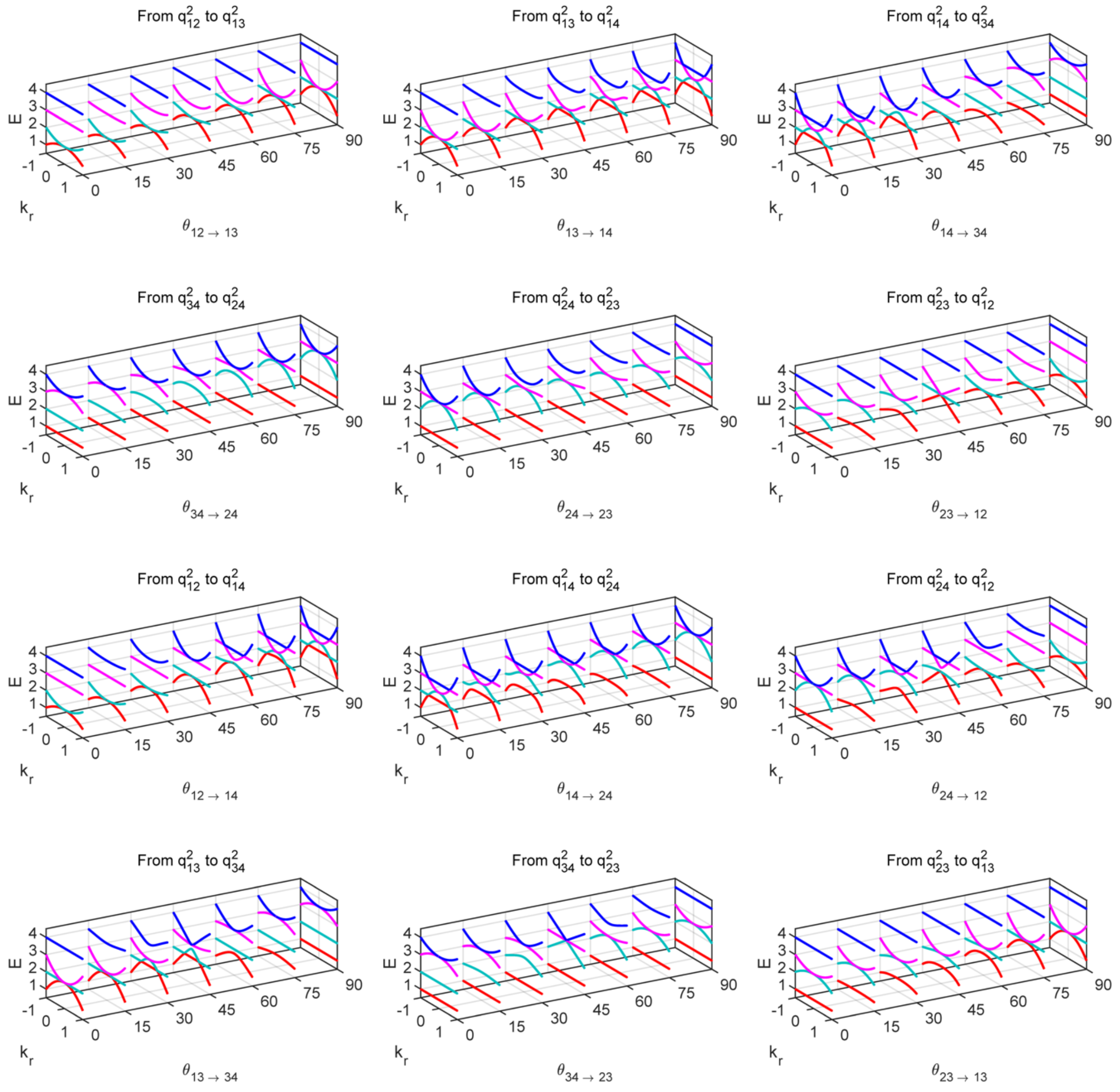

Figure S12. The evolution of radial cuts $E(k_r)$ of the extended two-dimensional bands between different factorizations of charge $-1$. The point degeneracies at $k_r = 0$ can be topologically related to the edge states of the 1D systems shown in Fig. S10. The other degeneracies ($k_r \neq 0$) are accidental without topological meaning.

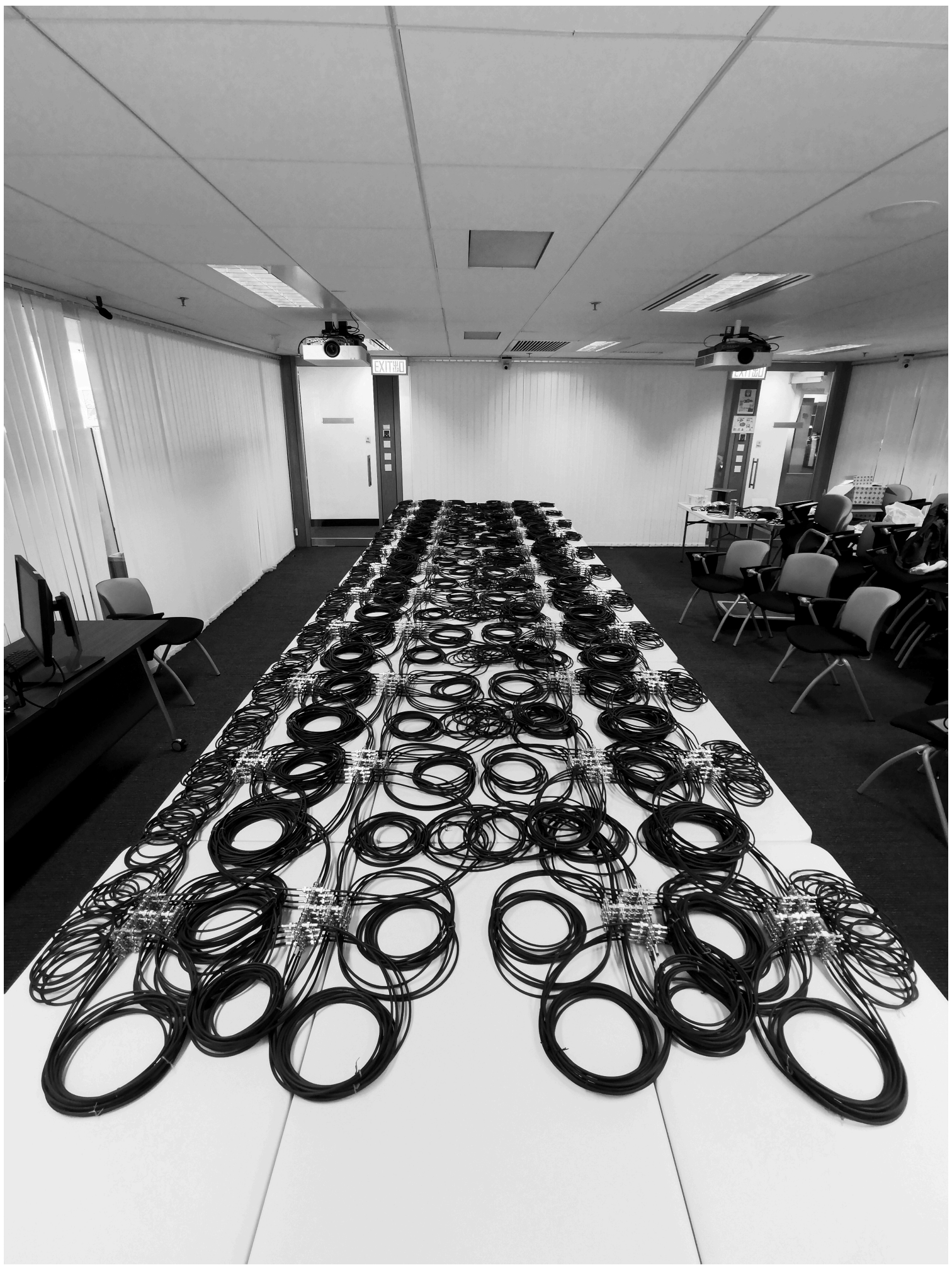

Figure S13. Transmission line network constructed for charges $\pm q_{14}$, where around 880 coaxial cables are used.

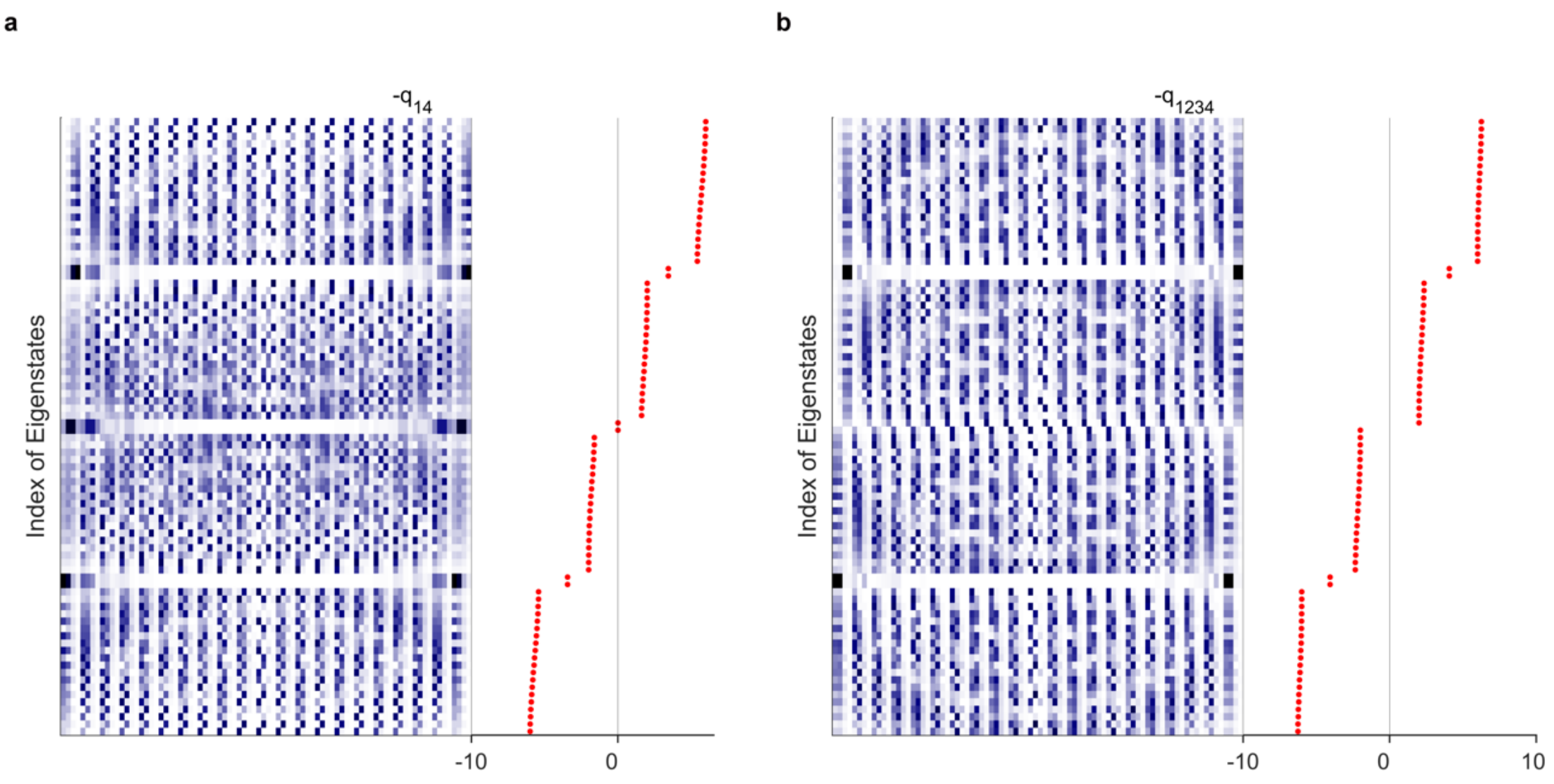

Figure S14. Distribution of hard boundary edge states for charges $\pm q_{14}$(a) and $-q_{1234} = -q_{12}q_{34}$(b). Detailed parameters are listed in Table S5.

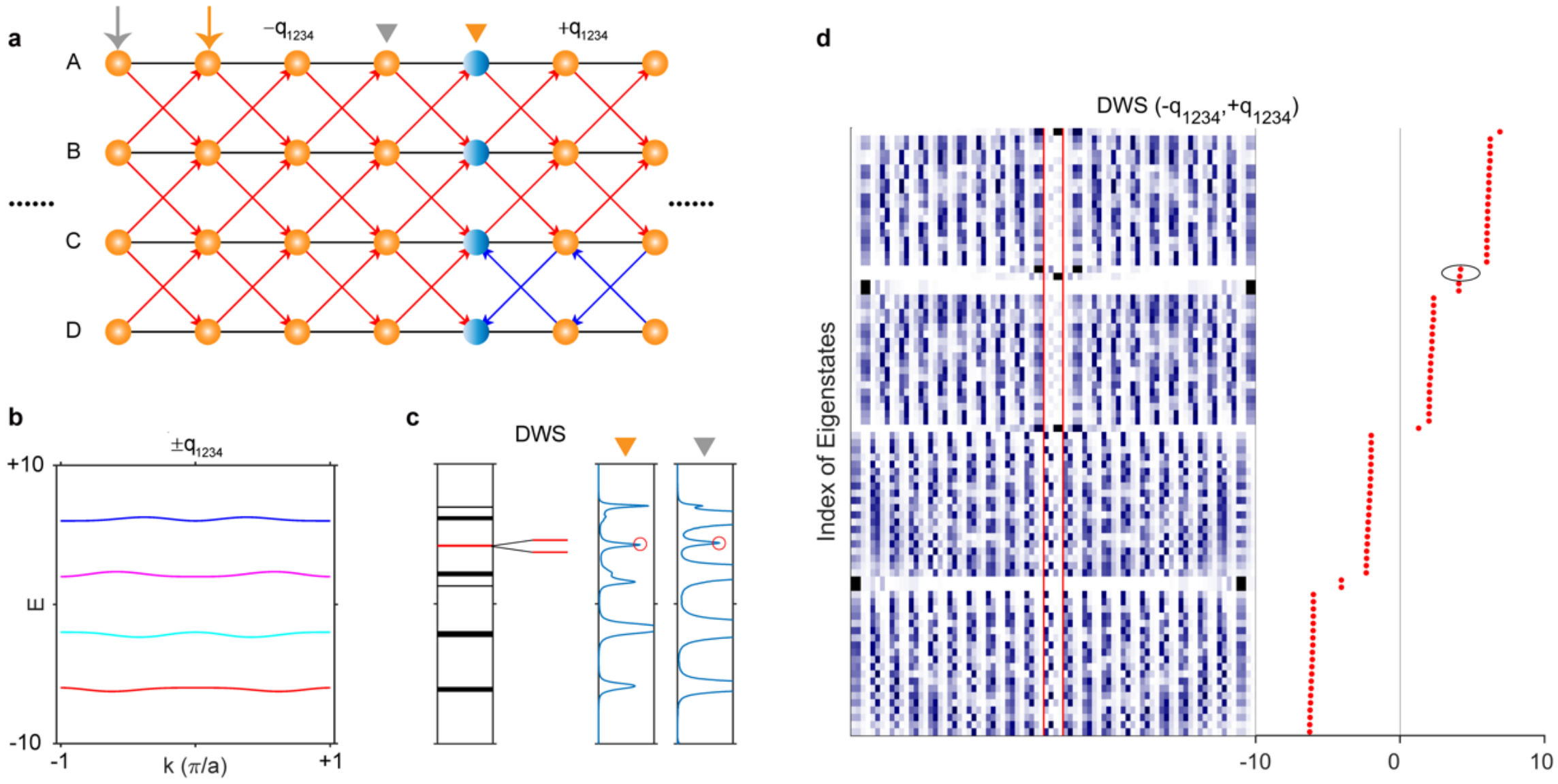

Figure S15. Construction of domain-wall (blue spheres in panel a) and distribution of domain-wall states (DWS) indicated by the black ellipse in panel (d). Panels (a) and (c) are copied from the main text (Fig. 4) for comparison purpose. b, Bulk states of $\pm q_{1234}$, they are overlapped. Detailed parameters are listed in Table S5 (see the column of $-q_{12}q_{34}$). For charge $q_{1234}$, we set $w_{CD} = -1$. We say that one domain-wall state locating in the second bandgap and another

one with energy beyond the bulk spectrum are induced by the domain-wall construction, and thus topologically trivial.

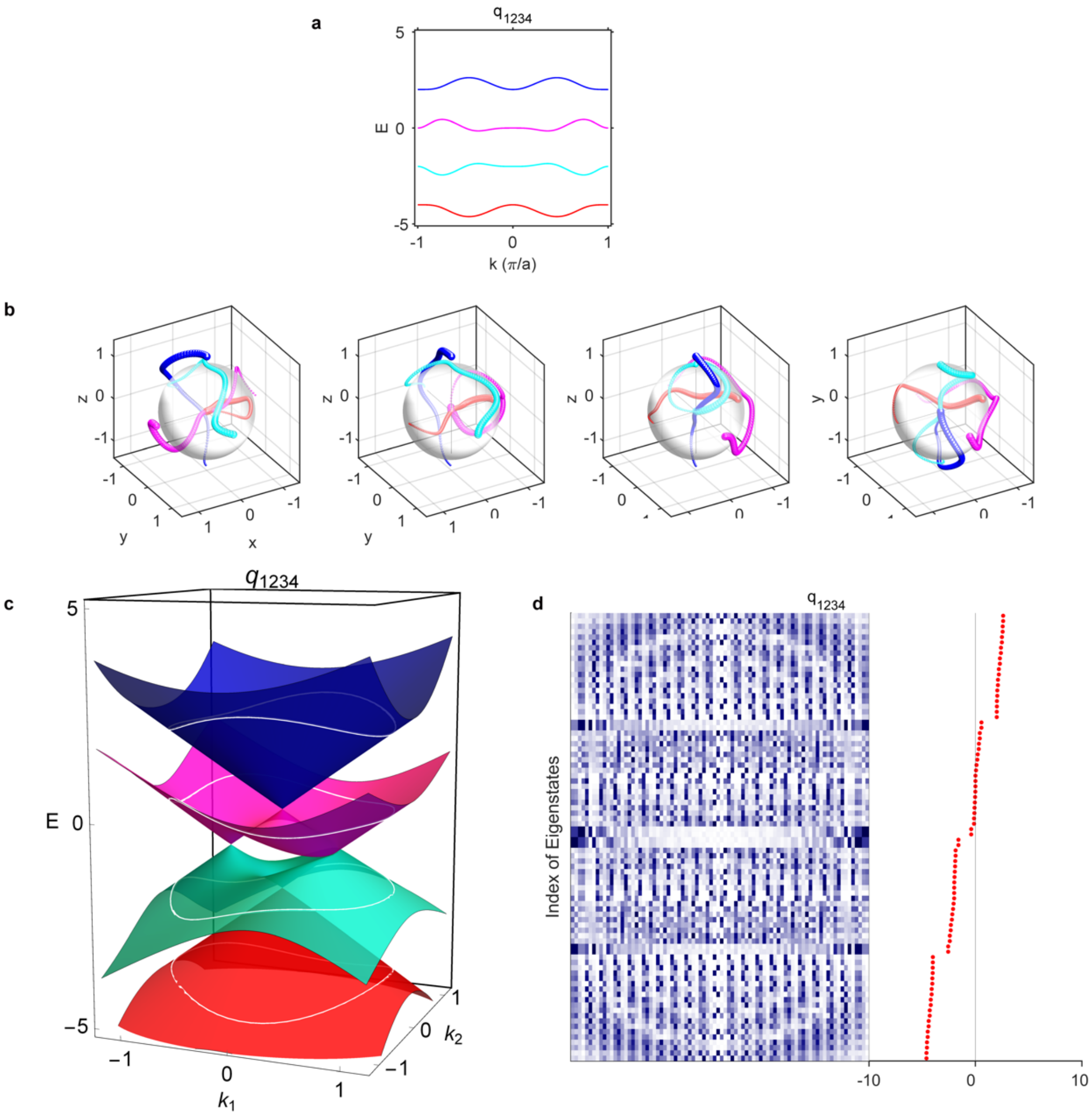

Figure S16. The case of $q_{1234} = -q_{13}q_{24}$. a, Bulk states. b, Trajectories of four eigenstates as wavevector runs across the first Brillouin zone ($k = -\pi \to \pi$). c, The extended energy bands on a 2D plane, where the white circles indicate the corresponding 1D energy bands. d, Distribution of hard boundary edge states. The first, second, third and fourth bands are coloured as red, cyan, magenta and blue, respectively. There is one linear Dirac cone between the first/third and second/fourth bands, and two linear Dirac cones between the second and third bands. Each linear Dirac cone implies one corresponding edge state per edge.

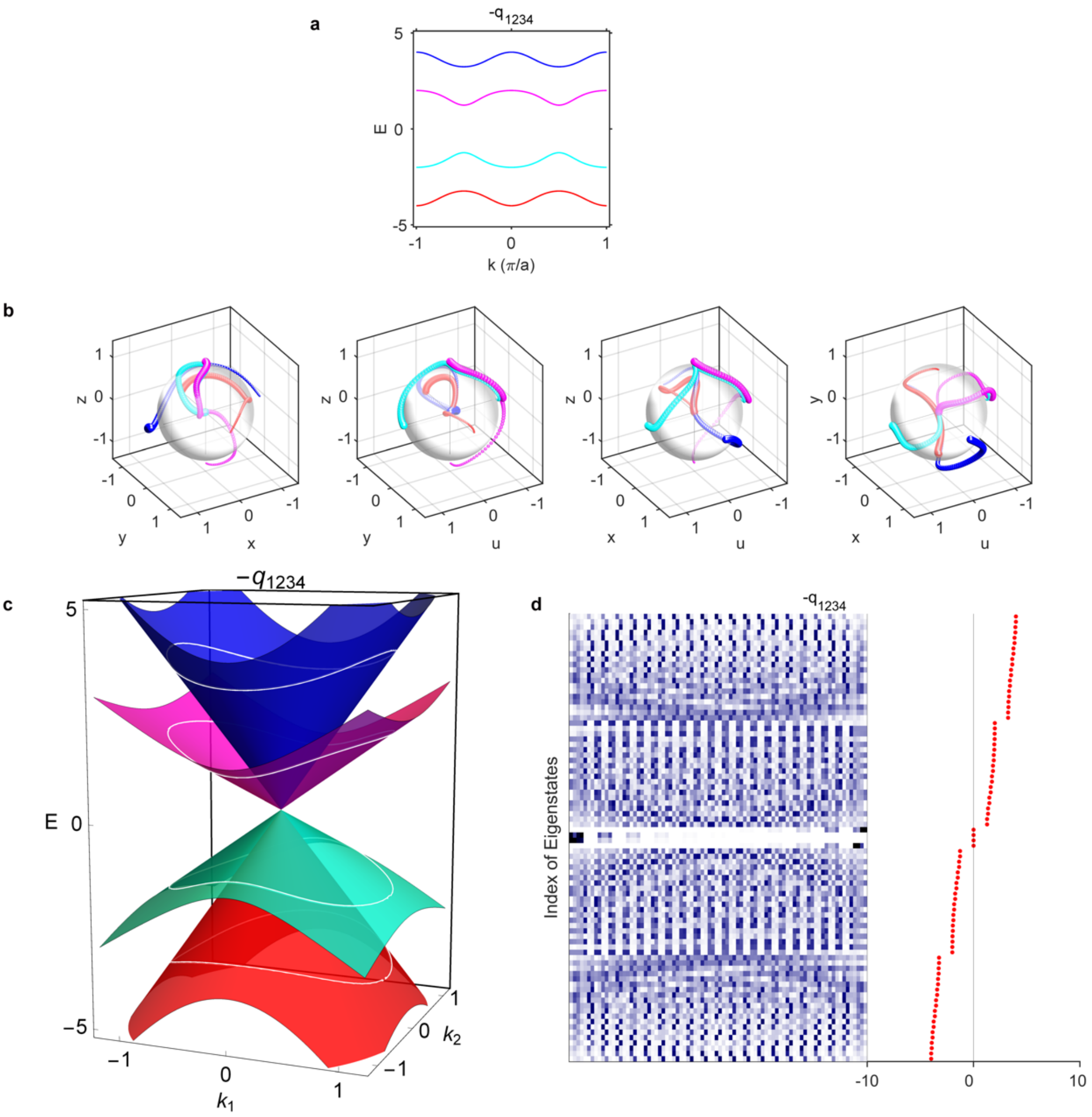

Figure S17. The case of $-q_{1234} = -q_{14}q_{23}$. a, Bulk states. b, Trajectories of four eigenstates as wavevector runs across the first Brillouin zone ($k = -\pi \rightarrow \pi$). c, The extended energy bands on a 2D plane, where the white circles indicate the corresponding 1D energy bands. d, Distribution of hard boundary edge states. The first, second, third and fourth bands are coloured as red, cyan, magenta and blue, respectively. There is one four-fold linear Dirac cone between the four bands, which implies two edge states per edge.

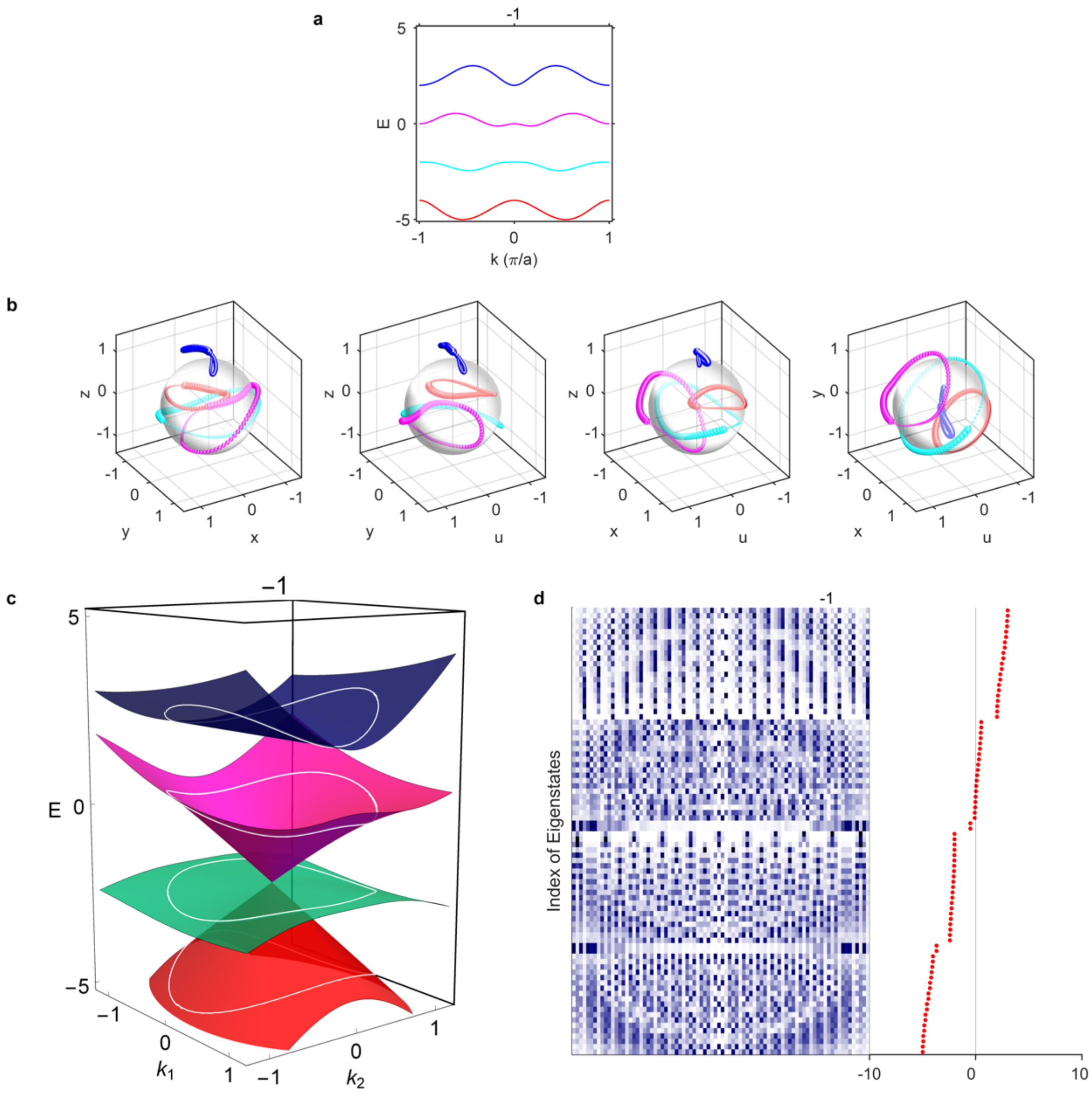

Figure S18. The case of charge $-1$. a, Bulk states. b, Trajectories of four eigenstates as wavevector runs across the first Brillouin zone ($k = -\pi \rightarrow \pi$). c, The extended energy bands on a 2D plane, where the white circles indicate the corresponding 1D energy bands. d, Distribution of hard boundary edge states. The first, second, third and fourth bands are coloured as red, cyan, magenta and blue, respectively. There is one triple linear degeneracy constructed by the lower three bands, which implies that for the first and second bandgaps each supports one edge state per edge, being similar to some cases of charge $-1$ in three-band models[22].